\documentclass[%
 reprint,
 superscriptaddress,
 amsmath,amssymb,
 aps,
 prl,
]{revtex4-2}
\usepackage{graphicx}
\usepackage{dcolumn}
\usepackage{bm}
\usepackage{mathtools}
\usepackage{float}
\usepackage[usenames,dvipsnames]{color}
\usepackage[colorlinks=true, citecolor=blue, linkcolor=blue, urlcolor=blue]{hyperref}
\usepackage{xr-hyper}
\usepackage{bm}
\usepackage{cleveref} 
\usepackage[nolist]{acronym}

\usepackage[version=4]{mhchem} 
\usepackage{chemfig} 
\setchemfig{atom sep=20pt} 

\usepackage{physics}

\usepackage[table]{xcolor} 
\usepackage{booktabs}      
\usepackage{multirow}      
\usepackage{array}         
\usepackage{tabularray}
\usepackage{nicematrix}

\definecolor{lightgray}{gray}{0.9}
\usepackage{tcolorbox}

\tcbuselibrary{breakable}

\begin{document}
\title{Fluctuation engineering in cavity quantum materials}

\newcommand{\affiliationRWTH}{
Institut f\"ur Theorie der Statistischen Physik, RWTH Aachen University and JARA-Fundamentals of Future Information Technology, 52056 Aachen, Germany
}

\newcommand{\affiliationMPSD}{
Max Planck Institute for the Structure and Dynamics of Matter,
Center for Free-Electron Laser Science (CFEL),
Luruper Chaussee 149, 22761 Hamburg, Germany
}
\newcommand{\affiliationPKS}{
Max Planck Institute for the Physics of Complex Systems,
Nöthnitzer Stra\ss e 38, 01187 Dresden, Germany
}
\newcommand{\affiliationBremen}{
Institute for Theoretical Physics and Bremen Center for Computational Materials Science,
University of Bremen, 28359 Bremen, Germany
}

\newcommand{\affiliationColumbia}{
Department of Physics, Columbia University, New York, NY, USA
}

\newcommand{\affiliationETH}{
Institute for Quantum Electronics, ETH Z\"{u}rich, Z\"{u}rich 8093, Switzerland}

\newcommand{\affiliationQC}{
Quantum Center, ETH Z\"{u}rich, Z\"{u}rich 8093, Switzerland}

\newcommand{\affiliationTrieste}{
Dipartimento di Fisica, Università degli Studi di Trieste, Trieste I-34127, Italy}

\newcommand{\affiliationErlangen}{
Department of Physics, University of Erlangen-Nürnberg, 91058 Erlangen, Germany}

\newcommand{\affiliationHamburg}{
Institute of Theoretical Physics, University of Hamburg, Notkestrasse 9, 22607 Hamburg, Germany}

\newcommand{\affiliationCUI}{The Hamburg Centre for Ultrafast Imaging, Hamburg, Germany}

\newcommand{\affiliationCCQ}{
Center for Computational Quantum Physics (CCQ) and Initiative for Computational Catalysis (ICC), The Flatiron Institute, 162 Fifth avenue, New York, NY 10010, USA}

\newcommand{\affiliationPenn}{
Department of Physics and Astronomy, University of Pennsylvania, Philadelphia, PA 19104, USA}

\newcommand{\affiliationUMD}{
Joint Quantum Institute, University of Maryland, College Park, MD 20742, USA
}

\newcommand{\affiliationKITP}{
Kavli Institute for Theoretical Physics, Santa Barbara, California 93106, USA
}

\author{Hope M.~Bretscher}
\email[Correspondence to: ]{hope.bretscher@mpsd.mpg.de}
\affiliation{\affiliationMPSD}
\affiliation{\affiliationColumbia}

\author{Lorenzo Graziotto}
\affiliation{\affiliationETH}
\affiliation{\affiliationQC}

\author{Marios H.~Michael}
\affiliation{\affiliationPKS}
\affiliation{\affiliationMPSD}

\author{Angela Montanaro}
\affiliation{\affiliationErlangen}

\author{I-Te Lu}
\affiliation{\affiliationMPSD}

\author{Andrey Grankin}
\affiliation{\affiliationUMD}

\author{James W.~McIver}
\affiliation{\affiliationMPSD}
\affiliation{\affiliationColumbia}

\author{J\'er\^ome Faist}
\affiliation{\affiliationETH}
\affiliation{\affiliationQC}

\author{Daniele Fausti}
\affiliation{\affiliationTrieste}
\affiliation{\affiliationErlangen}

\author{Martin Eckstein}
\affiliation{\affiliationHamburg}
\affiliation{\affiliationCUI}

\author{Michael Ruggenthaler}
\affiliation{\affiliationMPSD}

\author{Angel Rubio}
\affiliation{\affiliationMPSD}
\affiliation{\affiliationCCQ}

\author{D.N. Basov}
\affiliation{\affiliationColumbia}

\author{Mohammad Hafezi}
\affiliation{\affiliationUMD}
\affiliation{\affiliationKITP}

\author{Martin Claassen}
\affiliation{\affiliationPenn}
\affiliation{\affiliationKITP}

\author{Dante M.~Kennes}
\email[Correspondence to: ]{dante.kennes@rwth-aachen.de}
\affiliation{\affiliationRWTH}
\affiliation{\affiliationMPSD}

\author{Michael A.~Sentef}
\email[Correspondence to: ]{sentef@uni-bremen.de}
\affiliation{\affiliationBremen}
\affiliation{\affiliationMPSD}
\affiliation{\affiliationKITP}

\date{\today}

\begin{abstract}
 Coupling tailored electromagnetic fluctuations to materials provides a resource for controlling correlated quantum matter. By structuring the frequency, spatial, and modal distribution of fluctuations through a new generation of cavity quantum materials, vacuum and thermal spectra can shift phase boundaries and stabilize or suppress orders. This review organizes the field around a fluctuation-focused perspective, surveying a practical design toolbox and recent milestones, and outlining theory–experiment challenges in realistic, multimode, beyond-long-wavelength regimes. We highlight photonic observables   and map opportunities for equilibrium and driven control across superconducting, magnetic, moiré, and topological platforms. 
\end{abstract}

\maketitle

\begin{acronym}
\acro{EM}[EM]{electromagnetic}
\acro{QED}[QED]{quantum electrodynamics}
\acro{PF}[PF]{Pauli-Fierz}
\acro{DFT}[DFT]{density-functional theory}
\acro{QEDFT}[QEDFT]{quantum-electrodynamical density-functional theory}
\end{acronym}


\begin{figure*}
    
    \centering
    \includegraphics[width=\linewidth]{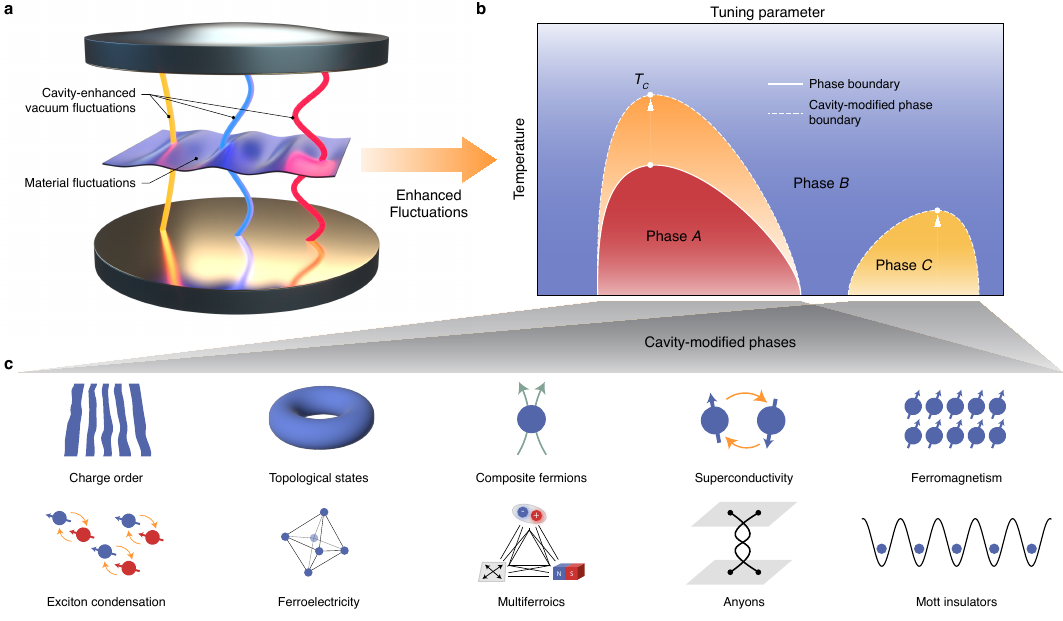}
    \caption{\textbf{Reshaping the phase diagram} $\vert$  (\textbf{a}) A quantum material with spatial and temporal fluctuations is coupled to fluctuations of the electromagnetic field engineered by a Fabry-Pérot cavity. (\textbf{b}) The coupling of material degrees of freedom to cavity-engineered fluctuations can transform the ground state phase diagram of the enclosed material, modifying transition temperatures or even resulting in the appearance of novel phases not accessible in the free-space environment. (\textbf{c}) Selection of possible fluctuation-engineered phases which have been experimentally realized or theoretically predicted. }
    \label{fig:simplefig}
\end{figure*}

Macroscopic properties of quantum materials are intimately tied to spatial and temporal correlations of fluctuating quasiparticles. Charge, spin, orbital, and lattice fluctuations—shaped further by quantum geometry, topology, and dimensionality—give rise to a complex landscape of competing orders~\cite{keimer2017quantum}. Close to phase boundaries, correlation lengths diverge and materials display extreme sensitivity, where even small perturbations can drive dramatic macroscopic changes. This inherent susceptibility to fluctuations represents both a central challenge and a unique opportunity: by learning how to manipulate these fluctuations, one may design and stabilize entirely new quantum phases.

Traditional approaches to alter microscopic interactions and explore the resulting phases rely on static tuning parameters such as doping, pressure, magnetic field, temperature, or static dielectric environment\cite{raja2017coulomb}.  More recently, ultrafast laser pulses have been used as a dynamical control knob realizing novel nonthermal quantum phases and engineering effective couplings via a semiclassical driving force~\cite{de_la_torre_colloquium_2021,basov_towards_2017}. However, the need for strong lasers, and practical challenges such as heating, decoherence, short-lived states, make it challenging to study or exploit these phases for applications.

The key role of fluctuations in quantum materials suggests an alternative strategy to understand and wield macroscopic quantum phenomena. In the past few years, the field of cavity quantum materials has opened a new route to engineer material degrees of freedom by hybridizing them with enhanced electromagnetic field modes in a cavity (Fig.~\ref{fig:simplefig}). Embedding a material in a cavity couples spatiotemporal fluctuations of light and matter, altering excitation spectra, mediating new interactions, or modifying electron localization, all of which can tip the balance between competing orders and stabilize different ground states. This offers new pathways to design and sustain quantum phases in situ, where material properties depend not only on atomic composition but also on photonic or polaritonic surroundings~\cite{schlawin_cavity_2022,hubener2021engineering,garcia-vidal_manipulating_2021}. Precision engineering of material correlations may also help disentangle the relevant microscopic degrees of freedom of strongly correlated materials, and provide novel observables through cavity quantum electrodynamical probes of material correlations.

Theory has outlined many opportunities for fluctuation control, for instance in superconductors~\cite{sentef.ruggenthaler.ea_2018, schlawin.cavalleri.ea_2019,curtis2019cavity,andolina_amperean_2024}, quantum magnets~\cite{QSL_Natcom2021,masuki.ashida_2024,Bostrom_Ferromagnetism}, 
excitonic systems~\cite{superradiant_Mazza}, quantum Hall and topological platforms~\cite{rokaj.penz.ea_2022,masuki.ashida_2023,CiutiTopology20241D}, or ferroelectrics~\cite{LatiniSTO2021,curtis2023local,ashida2020quantum}.
Advances in polaritonic chemistry have demonstrated that tailored electromagnetic environments can modify chemical reaction rates~\cite{garcia-vidal_manipulating_2021,ebbesen2016hybrid}, 
whereas developments in cavity fabrication have provided a toolbox with which to pattern fluctuations. With these in hand, experimental demonstrations indicate that cavity-engineered fluctuations, both quantum vacuum fluctuations and thermal photons, can indeed alter both excitation spectra and macroscopic responses ~\cite{appugliese2022breakdown,enknergraziottofraction,graziotto2025cavity,jarc_fausti_2023,keren2025cavity,thomas2021large}.

Yet, a gap remains between theoretical predictions, often for simplified models, and experimental realizations. Bridging this divide requires closer integration of experimental capabilities, theoretical modeling, and computational approaches. Open questions include: How precisely can fluctuations be tuned and characterized in correlated materials embedded in cavities? Which degrees of freedom are essential in theoretical descriptions, and which can be simplified? What are the clearest experimental signatures of cavity-modified ground or excited states? And how can iterative theory–experiment feedback accelerate progress?

\onecolumngrid
\vspace{2mm}
\begin{tcolorbox}[breakable,
  colframe=blue!40!black,
  colback=blue!8,
  coltitle=white,
  title=\Large Bridging cavity QED and quantum materials,
  fonttitle=\bfseries,
  left=10pt,right=10pt,top=10pt,bottom=10pt,
  width=\textwidth]

\quad A natural entry point to cavity quantum materials is cavity quantum electrodynamics (cQED), where a two-level atom interacts with a single-mode cavity (see \hyperref[theorybox]{Theory Overview Box}~(A)). Such quantum-optical systems are characterized by the light–matter coupling strength $g$~\cite{frisk_kockum_ultrastrong_2019,forn2019ultrastrong}. In the weak-coupling regime, dissipation (defined by the linewidth of the atom, $\gamma$, or cavity $\kappa$) dominates  ($\gamma/g$ or $\kappa/g <1$), preventing coherent exchange between subsystems, but still resulting in marked macroscopic changes, like Purcell enhancement of emission. Strong coupling ($g/\gamma > 1$) phenomena, like Rabi oscillations and normal-mode splitting, are shaped by the coherent exchange of energy between the cavity and atom. In the ultrastrong regime ($g/\omega \gtrsim 0.1$), counter-rotating terms and vacuum fluctuations become relevant, and the distinction between light and matter begins to blur~\cite{li2018vacuum}. For a detailed discussion, see reviews~\cite{frisk_kockum_ultrastrong_2019,forn2019ultrastrong}.

\quad For cavity quantum materials, these classifications are less central. Quantum materials host many coupled degrees of freedom, not reducible to single excitation energies and linewidths. Even modest cavity–matter interactions can alter the fluctuation spectrum, which, through anti-resonant contributions, can result in a new ground state, reshaping entire phase diagrams. Hybridization with the cavity can, for instance, enhance localization by dressing electrons with photon vacuum fluctuations and vice versa. Such self-consistent dressing can tip the balance between competing orders. More generally, the cavity modifies the electromagnetic environment and acts as a tailored reservoir: its spatiotemporal field fluctuations couple to material fluctuations, stabilizing or destabilizing ordered states. This motivates the search for new metrics to quantify and classify cavity coupling which go beyond coupled two-level pictures~\cite{eckhardt2024surface}.

\end{tcolorbox}
\twocolumngrid

\section{Fluctuations in cavity quantum materials}

As light–matter interactions are intrinsically weak in free space, employing them to gain control over a material requires enhancing the electromagnetic field fluctuations. A Fabry–Pérot resonator achieves this through reflecting boundaries that confine photons so that they traverse the enclosed region many times, strongly enhancing their coupling to matter placed inside.

\begin{figure*}
    \centering
    \includegraphics[width=\textwidth]{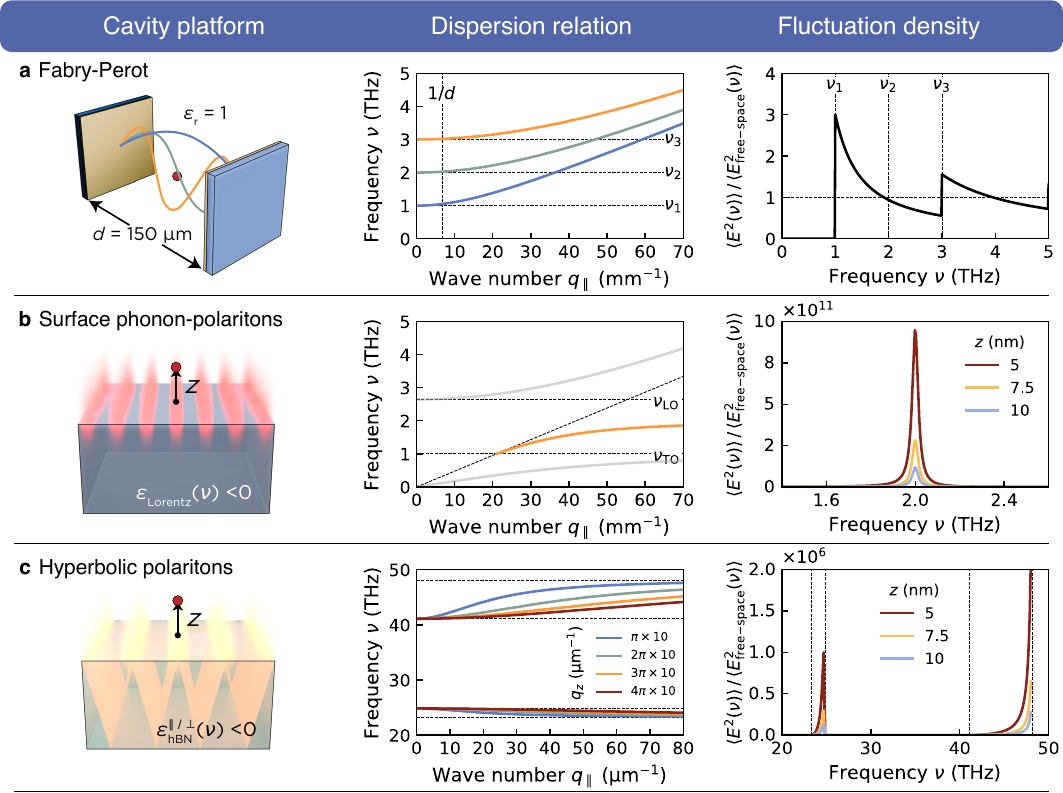}
    \caption{\textbf{Cavity engineered fluctuations} $\vert$ Pictorial representation of exemplary cavity platforms (left), their dispersion relation (center), and fluctuation density---calculated at the location indicated by the red dot marker in the left panel---reported as the variance of the electric field as a function of frequency, normalized to the free space value (right). \textbf{(a)} Fabry-Pérot cavity, with the three lowest modes. \textbf{(b)} Surface phonon polaritons, with dielectric function $\varepsilon_\mathrm{Lorentz}$ given by the Lorentz oscillator model with longitudinal and transverse optical phonon frequencies $\nu_\mathrm{LO}$ and $\nu_\mathrm{TO}$, respectively. \textbf{(c)} Hyperbolic polaritons, with dielectric function $\varepsilon_\mathrm{hBN}^{\parallel},\varepsilon_\mathrm{hBN}^{\perp}$ given by the Lorentz oscillator model for in-plane and out-of-plane optical phonons, respectively.}
    \label{fig:fig3}
\end{figure*}

The boundary conditions that confine real photons also reshape the spectrum of electric-field fluctuations inside a cavity (Fig.~\ref{fig:fig3}). Even at zero temperature the electromagnetic ground state is not static: Heisenberg's uncertainty principle forbids the simultaneous vanishing of electric and magnetic field energies. Hence the field exhibits zero-mean fluctuations but a finite frequency $\nu$-dependent variance $\langle \Delta E^2(\nu)\rangle$, the vacuum fluctuations (Fig.~\ref{fig:fig3}a).
Matter couples to these fluctuations through the exchange of ``virtual photons'', short-lived excitations that mediate interactions. At finite temperature, thermal radiation supplies a bath of real photons, adding occupation-dependent fluctuations in photon number $\langle \Delta n^2 \rangle$ and electric field $\langle \Delta E^2 \rangle$ that are governed by the photonic density of states. By restricting the modes sustained between parallel plates, a Fabry–Pérot cavity redistributes the electric-field fluctuations, enhancing them at certain frequencies while suppressing them at others (Fig.~\ref{fig:fig3}a).

\begin{figure*}
    \centering
    \includegraphics[width=\linewidth]{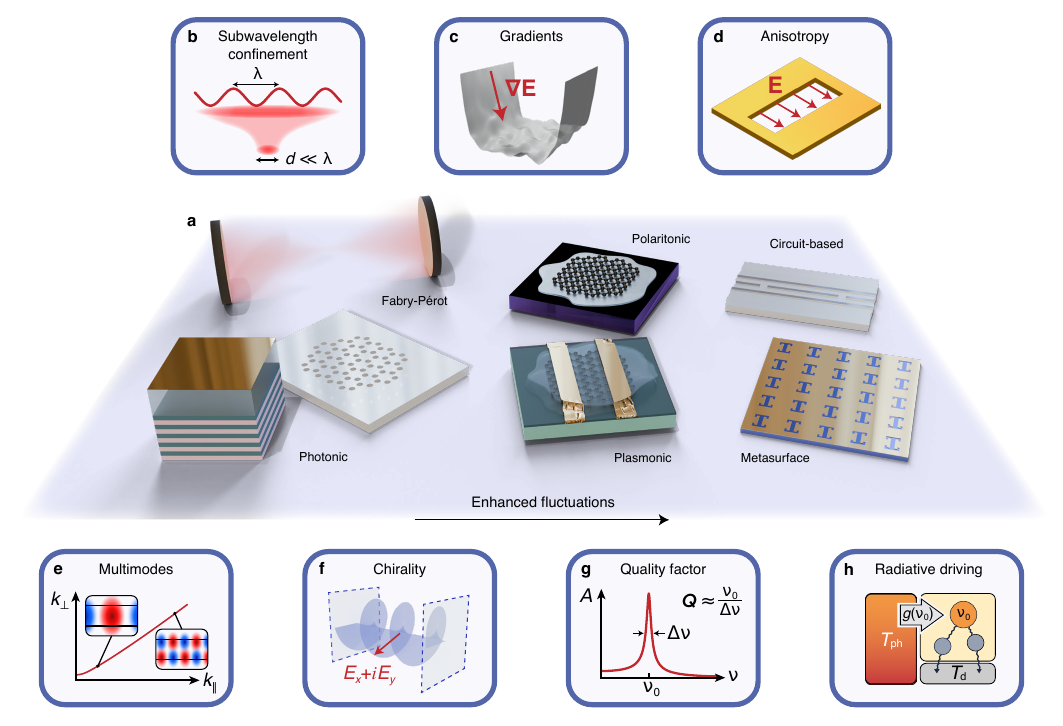}
    \caption{\textbf{Diverse platforms and design principles} $\vert$ \textbf{(a)} Cavity platforms, roughly arranged from left to right by increasing electromagnetic field confinement, resulting in enhanced fluctuations. \textbf{(b-h)} Tuning knobs explored by cavity engineering, whose behavior is elucidated in Sec.~\hyperref[sec:designprinciples]{The Cavity Design Toolbox}. $\lambda$, wavelength; $d$, length scale of spatial field confinement; $\nabla \vb{E}$, gradient of the electric field; $\vb{E}$, electric field; $k_\parallel, k_\perp$, mode wavevector components; $E_x, E_y$, electric field components; $A$, spectral function; $Q$, quality factor; $\nu_0$, mode central frequency; $\Delta \nu$, mode linewidth; $\nu$, frequency; $T_\mathrm{ph}$, photon bath temperature; $g(\nu_0)$, selective coupling between the cavity and the matter mode at frequency $\nu_0$; $T_\mathrm{d}$, dissipative bath temperature.}
    \label{fig:cavityPlatforms}
\end{figure*}

In cavity quantum materials, this reshaping of fluctuations is used to influence macroscopic properties of matter across widely disparate energy scales. Modern cavity architectures extend beyond the Fabry–Pérot paradigm and include split-ring resonators~\cite{enknergraziottofraction,graziotto2025cavity,appugliese2022breakdown}, nano-tip-based cavities~\cite{park2019tip,koo2023tunable}, photonic microcavities~\cite{tay2025multimode,zhang2018photonic}, and metasurface or metamaterial designs~\cite{sarkar2025sub,huang2023tunable,sortino2025atomic} (Fig.~\ref{fig:cavityPlatforms}). We use the term “cavity” to indicate any form of electromagnetic environment that hosts engineered bosonic modes coupled to quantum matter~\cite{herzig2024high,gogna2020self,kipp2024cavity,sternbach2020femtosecond}. This definition encompasses polaritonic platforms~\cite{basov_polariton_2021,basov2025polaritonic}, including phonon, plasmon, or exciton polaritons. Notably, polariton cavities introduce hybrid light-matter states even in the dark, where material excitations are dressed by virtual photons and vice versa. Surface phonon polaritons, for instance, sharply renormalize the fluctuation spectrum at discrete frequencies and within the distance set by their decay length (Fig.~\ref{fig:fig3}b)~\cite{eckhardt2024surface,eckhardt_theory_2024}; these collective modes can be viewed as effective quantum fields whose fluctuations reflect a feedback loop between the underlying material and its hybridization with light. Remarkably, many materials and heterostructures can even act as \emph{self-cavities}, where a large dielectric permittivity, layered geometry, and device or sample edges can sustain waveguide-like modes that hybridize with polaritons~\cite{dirnberger2023magneto,gogna2020self,kipp2024cavity,sternbach2020femtosecond}, underscoring how ubiquitous cavity-induced fluctuation effects have become in contemporary materials platforms.

\section{The Cavity Design Toolbox}\label{sec:designprinciples}

While it is typically important that the cavity design is tailored to match the characteristic scales of a quantum material and its phases of interest, relevant design levers for cavity engineering extend far beyond energy, length, and dimensional scale matching. 

\subsection{Subwavelength confinement}

Subwavelength confinement of electromagnetic fields can enhance vacuum fluctuations and local electric field amplitude by orders of magnitude relative to free space (see Fig.~\ref{fig:cavityPlatforms}b). This enhancement leads to an increased light–matter coupling strength $g$, e.g., of a dipole moment $\vb{d}$ to a field $\vb{E}$ ($g\propto \vb{d}\cdot\vb{E}$).

In the ultrastrong regime, higher-order terms in the light–matter Hamiltonian, which are negligible at weaker coupling, become relevant. These terms can mediate new effective interactions between material degrees of freedom and thus provide qualitatively new avenues of control~\cite{juraschek2021cavity,mornhinweg2021tailored,masuki.ashida_2023,bacciconi2025theory,arwas.ciuti_2023,ciuti2021cavity}. Experiments in this regime have observed dark-cavity modified electronic transport of cavity-coupled Landau polaritons~\cite{paravicini2019magneto}. Low-energy and subwavelength cavities are thus advantageous for attaining coupling regimes where non-perturbative effects become prominent. 

\subsection{Gradients and anisotropy}\label{toolboxgradients}

Another powerful design element are spatial gradients of the electromagnetic field (see Fig.~\ref{fig:cavityPlatforms}c). Unlike Fabry–Pérot cavities, which support nearly homogeneous modes, planar split-ring resonator structures~\cite{enknergraziottofraction,appugliese2022breakdown} or plasmonic cavities with patterned dielectric environments~\cite{kipp2024cavity} generate strongly inhomogeneous near fields. At the resonator edges, the abrupt change of field orientation produces gradients as large as $10^{8}$~V/m$^{2}$.

Spatial gradients play a central role in circumventing no-go theorems for the Dicke superradiant phase transition derived under the assumption of homogeneous fields~\cite{nataf2010no} and were theoretically shown to affect topological protection~\cite{rokaj2023weakened}. Strengthened fractional quantum Hall states observed in two-dimensional electron systems coupled to resonators (Fig.~\ref{fig:flagship}d-e) were attributed to electron–electron interactions within a Landau level, made possible by spatially inhomogeneous coupling~\cite{enknergraziottofraction}, whereas integer quantum Hall states were observed to be weakened by cavity-coupling, attributed to a form of photonic disorder~\cite{ciuti2021cavity}. 

 Large gradients are also expected to couple to excitations without a dipole moment and play a role in itinerant electron systems where the dipole approximation breaks down. Multi-polar interactions become relevant, and net angular momentum can be imparted~\cite{bacciconi2025theory,session2025optical}. This implies that cavities can modify quantum materials even via multi-polar or Raman-active modes, extending their influence well beyond conventional dipole-driven processes. 

The spatial shaping of vacuum fluctuations is particularly important for anisotropic electronic phases. A prime example is the quantum Hall stripe phase, a charge density wave emerging at ultralow temperatures in compressible regimes~\cite{koulakov1996charge}. Recent experiments demonstrate that anisotropic fluctuations inside a slot-antenna resonator—with modes polarized orthogonal to its long axis (Fig.\ref{fig:cavityPlatforms}d)—can align the stripe orientation orthogonal to the cavity field~\cite{graziotto2025cavity}. To minimize the free energy, the stripes orient their high-conductivity axis opposite to the strongest fluctuations, corresponding to the Casimir energy minimum in an anisotropic cavity~\cite{graziotto2025cavity}.
This results in a fifty-fold suppression of the longitudinal resistivity, realizing cavity-enhanced transport. More generally, the coupling of anisotropic fluctuations to quantum phases highlights cavity anisotropy as a control knob for patterning electronic dispersion, akin to moiré engineering.


\subsection{Multi-modal cavities}

An interesting playground is provided by interactions with a continuum of modes. Multi-modal polaritonic cavities---such as those based on hyperbolic polaritons---are highly non-dispersive (Fig.~\ref{fig:fig3}c), enhancing fluctuations by coupling across a broad range of modes (see Fig.~\ref{fig:cavityPlatforms}e and \hyperref[theorybox]{Theory Overview Box}~(B)). Accessing these multi-modal fluctuations requires positioning a quantum material within the evanescent decay length ($\sim$ nm--\textmu m) of the cavity field. When proximitized within a length scale commensurate to that of the lattice or the relevant Fermi wavevector ($k_F$), the cavity can induce $q\neq 0$ interactions. 

Theoretical studies predict that near-field cavities composed of surface phonon polaritons in SrTiO$_3$ or hyperbolic phonon polaritons in thin-film hexagonal boron nitride (hBN) can enhance cavity-mediated effects and exhibit qualitatively different phenomena~\cite{keren2025cavity,Ashida2021}. For example, multi-modal coupling, such as to higher-momenta modes and fluctuations of a surface phonon-polariton, can lead to ultrastrong, polaron-like renormalization of the effective mass~\cite{eckhardt2024surface}, experimentally observed in Ref.~\cite{keller2017few}. 

\subsection{Mode polarization}
Another promising direction is to design the polarization of the cavity mode, such as the helicity and orientation of vacuum fluctuations. While related to cavity anisotropy, polarization control is more general: it targets the internal symmetry of the light field itself rather than spatial inhomogeneity of its intensity profile, and also enables coupling to anisotropies built into the material. Recent first-principles work predicts universal linear polarization-driven charge localization in 2D van der Waals systems that tunes band gaps, valley energies, and even interlayer spacing, enabling the control of ferroelectricity, nonlinear Hall responses, and tailored optical spectra at equilibrium~\cite{liu2025modifying}. Extending this idea to cavities with nontrivial polarization textures, such as chiral cavities (see Fig.~\ref{fig:cavityPlatforms}f), opens further possibilities for controlling broken-symmetry phases~\cite{hubener2021engineering} (Sec.~\hyperref[sec:outlookphases]{Cavity topology}), or amplifying chiral light-matter interactions~\cite{andberger2024terahertz}).

\subsection{Quality factor}
 In conventional cavity QED of isolated two-level systems, coherent energy exchange between matter and photons requires a high $Q$, defined as $Q=\nu/\gamma$, with $\nu_c$ the cavity frequency and $\gamma$ its linewidth (see Fig.~\ref{fig:cavityPlatforms}g). Optical and superconducting cavities in the canonical \textit{strong-coupling} regime ($Q>1$) often operate with $Q>10^4$ to ensure that coherent Rabi oscillations can be resolved~\cite{blais2021circuit}.

However, a high quality factor is not necessary and may even be undesirable for cavity quantum material engineering, where the goal is to enhance fluctuations. Experimental realizations relevant to Sec.~\hyperref[sec:exp]{Flagship experiments} typically involve quality factors on the order of $\omega/\gamma \sim 5$~\cite{enknergraziottofraction,jarc_fausti_2023,appugliese2022breakdown,thomas2021large}. Theory indicates that lowering $Q$ mainly alters the quantitative strength of cavity-induced effects, without changing their qualitative nature~\cite{virtual2017}. Lower quality factors can be advantageous, as a broadened spectral range of enhanced fluctuations can couple to a continuum of material excitations rather than to a single sharply defined resonance. In this sense, lossy cavities can act as broadband mediators, extending cavity control to situations where the relevant material modes are distributed in energy. Moreover, the interplay of dissipation and coupling strength provides an additional design knob: balancing confinement, linewidth, and mode structure allows tailoring whether the cavity acts more like a sharp resonant filter or a broadband fluctuation bath.

\subsection{Radiative driving}
\begin{figure*}
    \centering
    \includegraphics[width=.5\textwidth]{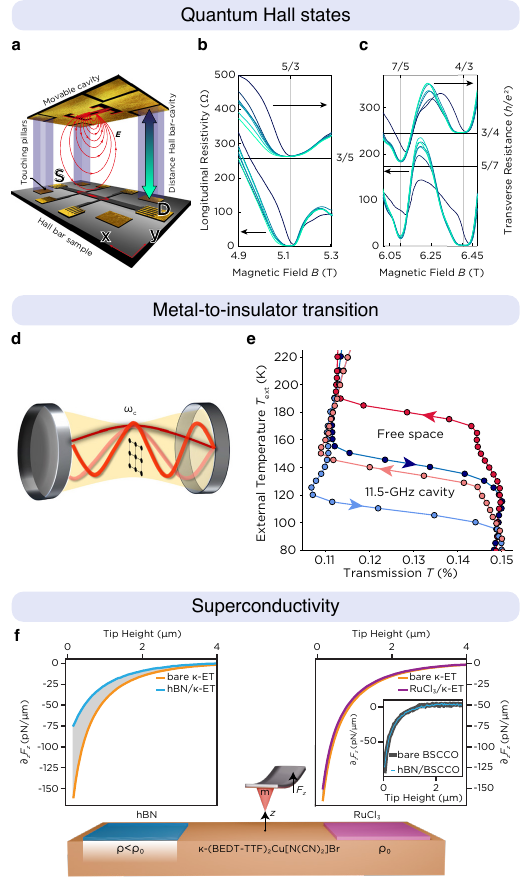}
    \caption{\textbf{Flagship cavity quantum materials experiments} $\vert$ \textbf{(a)} Metasurface cavity coupled to a two-dimensional electron system with tunable light-matter interaction strength controlled by the distance between the Hall bar sample and the movable cavity. The red lines indicate the cavity vacuum electric field profile. S, electric current source; D, drain. \textbf{(b)} Some fractions belonging to 1/3 Jain's family (here 5/3) are rendered more robust by increasing the light-matter coupling, i.e.\ reducing the cavity-Hall bar distance (from darker to lighter curves, according to the color bar shown in panel \textbf{a}). This is manifest from the modification of the longitudinal resistivity (left axis) and transverse resistance (right axis), which become flatter and span a broader magnetic field range (bottom axis). \textbf{(c)} Same as panel \textbf{b} but for fractions 7/5 and 4/3, still part of the 1/3 Jain's family. \textbf{(d)} A Fabry-Pérot cavity with fundamental angular frequency $\omega_c$ controls the radiative heat load on \ce{TaS2}. \textbf{(e)} The cavity (blue curve) shifts the transition temperature by 30~K with respect to free space (red curve), as demonstrated by the hysteresis cycle as a function of external temperature of the THz transmission (shown here integrated between 0.2--1.5~THz). \textbf{(f)} The polaritonic cavity set up at the interface between the $\kappa$-ET molecular superconductor (orange block) and the exfoliated hBN crystal (light blue square, left side of the figure) alters the superfluid density of $\kappa$-ET, as shown by the increase of the Meissner force derivative $\partial_z F_z$ (left axis), measured with magnetic force microscopy as a function of tip height (top axis). This behavior is specific to the hBN/$\kappa$-ET interface, as shown in the right side of the figure, where \ce{RuCl3}/$\kappa$-ET and hBN/BSCCO interfaces are studied. $z$, tip height; m, magnetic tip; $F_\mathrm{z}$, Meissner force on the magnetic tip; $\rho_0$, bare $\kappa$-ET superfluid density. \\[0.5em]
    Panels \textbf{a}-\textbf{c} adapted from Ref.~\cite{enknergraziottofraction} under a Creative Commons License CC-BY 4.0. Panels \textbf{d}-\textbf{e} adapted with permission from Ref.~\cite{jarc_fausti_2023}. Panel \textbf{f} adapted from Ref~\cite{keren2025cavity} under a Creative Commons License CC-BY 4.0.}
    \label{fig:flagship}
\end{figure*}

The driven-dissipative nature of a cavity can itself be exploited as a design parameter~\cite{inoue2015realization,cho2023directional}. The complex thermodynamic landscape of quantum materials can experience a modified effective dissipative environment when placed inside a cavity. For instance, in a Fabry–Pérot geometry, the cavity can act as a spectral filter, shaping the black-body radiation spectrum that reaches the quantum material and selectively loading specific modes with energy provided by the external thermal photonic bath (see Fig.~\ref{fig:cavityPlatforms}h). The impact of this mechanism has been highlighted in recent experiments on charge-density-wave TaS$_2$, where cavity electrodynamics were invoked to explain an anomalously large and non-monotonic radiative heat load~\cite{jarc_fausti_2023,jarc_fausti_2024} (Fig~\ref{fig:flagship}d-e). More generally, engineering the spectral and spatial distribution of photons in cavities can either inject energy into targeted material modes or, conversely, shield them from environmental disturbances. This controlled use of a natural thermal photonic bath to engineer thermal or non-thermal energy distribution inside the light-matter assembly thus provides a complementary pathway to cavity design, expanding the toolbox for manipulating quantum matter~\cite{fassioli_fausti_2024, flores_piazza_2025}.

\section{Flagship experiments} \label{sec:exp}
A selection of landmark experiments demonstrates the macroscopic consequences which can arise when quantum materials are embedded in tailored cavity environments that utilize the toolbox described above.

\subsection{Quantum Hall states}
Recent studies on quantum Hall phases have demonstrated the sensitivity of topological states to cavity vacuum fluctuations. Quantum Hall phases emerge in two-dimensional electron systems in a strong magnetic field at ultralow temperatures. They are characterized by dissipationless edge transport and precise quantization of the transverse conductivity to integer or fractional multiples of $e^2/h$. Their relevance for testing the influence of the electromagnetic environment stems from the large effective light-matter interaction length scales. In addition, the magnetic field suppresses the kinetic energy, highlighting the importance of the Coulomb interaction, whose form depends strongly on the electromagnetic surroundings. Metasurface structures with resonances in the 0.1--1~THz range are a suitable cavity choice for the planar geometry of the tens of micrometer-size samples~\cite{scalari2012ultrastrong}, with electromagnetic fields of wavelength $\lambda$ confined to subwavelength volumes $V\sim10^{-4}\lambda^3$, thus enhancing vacuum fluctuations to amplitudes of about {1 V/m} for the fundamental mode~\cite{paravicini2019magneto}. Recent experiments on integer quantum Hall phases have shown that electronic edge states are affected by such cavity vacuum fluctuations---i.e.\ they acquire a finite resistivity---which break their topological protection~\cite{appugliese2022breakdown} and suppress the single-particle energy gap~\cite{enknergraziottofraction}. For fractional phases, instead, the effect is the opposite: cavity vacuum fluctuations have been experimentally demonstrated to increase the transport gap of some fractions of intermediate fillings between $\nu = 1$ and $\nu = 2$ and belonging to the 1/3 Jain family by up to 50\%~\cite{enknergraziottofraction} (see Fig.~\ref{fig:flagship}a-c), due to a cavity-induced electron-electron attraction, which counteracts the long-range Coulomb repulsion and increases $T_c$~\cite{enknergraziottofraction}. Crucially, such interactions are thought to arise due to the high $10^8~\mathrm{V/m^2}$ gradients of the fluctuating field, as discussed in Sec.~\hyperref[sec:designprinciples]{The Cavity Design Toolbox}.

\subsection{Metal-to-insulator transition}
Experiments on the layered transition metal dichalcogenide 1T-\ce{TaS2} demonstrate that cavity control can be used to tune electronic phase transitions. This material exhibits multiple charge-ordered states stabilized by competing interactions. At around 180 K it undergoes a first-order, hysteretic transition from a nearly commensurate metallic charge-density-wave (NC-CDW) to an insulating commensurate CDW. Embedding 1T-\ce{TaS2} in a Fabry–Pérot cavity with tunable modes in the tens to hundreds of GHz range~\cite{jarc2022tunable} lowers this transition temperature by as much as 30 K (Fig.~\ref{fig:flagship}d-e). The shift is fully reversible: the system can be driven between metallic and insulating states simply by adjusting the cavity mirror spacing and alignment, with a distinctly non-monotonic dependence on the cavity resonance frequency~\cite{jarc_fausti_2023,jarc_fausti_2024}. This striking effect is attributed to the cavity-regulated injection of fluctuations from the external photon bath, highlighting how radiative dissipation engineering can be exploited to control correlated electron phases~\cite{fassioli_fausti_2024}.

\subsection{Superconductivity}
Superconductivity is another paradigmatic example of a phase that can be sensitive to changes in the environment. Unconventional superconductivity arises in families of layered materials with strong electronic correlations---most notably cuprates, iron pnictides, and organic superconductors~\cite{scalapino2012common,keimer2015quantum}. The layered organic salt $\kappa$-(BEDT-TTF)$_2$Cu[N(CN)$_2$]Br ($\kappa$-ET) is known to superconduct below $T_c=11.5$~K. In proximity to hBN, the superfluid density of $\kappa$-ET was reported to be sharply suppressed~\cite{keren2025cavity} (see Fig.~\ref{fig:flagship}f). hBN is a hyperbolic van der Waals material~\cite{dai2014tunable} with dielectric permittivities of opposite signs along different crystal axes. This hyperbolicity leads to infrared-active phonon polaritons with a sharply enhanced density of states, which appears near resonance with a 1470~cm$^{-1}$ \chemfig{C=C} phonon in $\kappa$-ET (see Fig.~\ref{fig:fig3}c). Control experiments suggest the resonant coupling between hyperbolic hBN polaritons and the \chemfig{C=C} phonon as the most likely origin for the cavity alteration of the superfluid density. As the superfluid density is the ground state property of a superconductor linked to the thermodynamics of the superconducting phase transition, this experiment~\cite{keren2025cavity} sets a precedent for cavity-altered thermodynamic properties.  Intriguingly, resonant laser driving of the same \chemfig{C=C} mode has been shown to lead to light-induced superconductivity~\cite{buzzi2020photomolecular}, raising the prospect that changes in the cavity design could enhance, rather than quench, superconductivity (see Sec.~\hyperref[sec:outlookphases]{Equilibrium control of established phases} for further discussions).

\section{Theoretical approaches}\label{sec:theory}

Theoretical approaches often make progress toward deliberate fluctuation engineering by inverting the problem: given a target material property, what is the most plausible mechanism and corresponding cavity design to modify it? Answering this question requires treating quantum and thermal fluctuations in both light and matter degrees of freedom, ideally on an equal footing. This is challenging as the ground and thermal states of a cavity quantum material cannot generally be described as a tensor product of quantized light and matter; the fluctuations of the emergent system are more than the sum of their parts. State-of-the-art theoretical approaches typically go beyond perturbation theory, either calculating the physical observables of the coupled light-matter system from first principles (Sec.~\hyperref[subsec:firstprinciples]{First-principles approach}), or investigating a restricted set of matter and light degrees of freedom to obtain the correlated wave functions (Sec.~\hyperref[subsec:effectivemodels]{Effective models}). 

\subsection{First-principles approach}\label{subsec:firstprinciples}

The basic theory that describes the non-perturbative interaction between the bare electrons, ions, and photons, can be deduced directly from special relativity~\cite{ruggenthaler.sidler.ea_2023}. The resulting Pauli--Fierz (PF) quantum field theory (see \hyperref[theorybox]{Theory Overview Box}~(F)) couples the bare electronic and ionic currents minimally to the full continuum of free-space photon modes~\cite{ruggenthaler.tancogne-dejean.ea_2018}. The PF Hamiltonian is formulated in the Coulomb gauge~\cite{ruggenthaler.sidler.ea_2023}, ensuring a clear separation between transverse and longitudinal fields in both free-space and cavity environments. However, this requires that the cavity material is also described microscopically. Solving practical problems requires approximations, as the full light-matter wave function depends on many degrees of freedom across length and energy scales.   

The PF is often reformulated in terms of collective variables that are computationally tractable and avoid gauge-dependent light-matter many-body wave functions. Quantum-electrodynamical density-functional theory (QEDFT), for instance, does so in terms of the three-dimensional charge-current densities and electromagnetic fields. This computational simplification requires approximating the quantum stress tensors to capture non-trivial quantum effects. These approximations are usually performed with the help of auxiliary systems, such as coupled Kohn-Sham and Maxwell equations, resulting in longitudinal (Coulombic) and transverse (photonic) exchange-correlation functionals~\cite{ruggenthaler.sidler.ea_2023}.

The multi-scale character of the problem makes further simplifications necessary. The long-wavelength approximation, which neglects momentum transfer from cavity photons, is often assumed, allowing QEDFT descriptions for the material inside the cavity to be combined with effective theories for cavity modes, such as macroscopic QED~\cite{svendsen2024ab}. Moreover, since a cavity modifies only a small part of the spectrum of the photonic density of states, one can subsume the vast majority of the continuum of modes in the observable, renormalized masses of the electrons and ions and keep only the difference to free-space~\cite{svendsen_2023}. 

\textit{Ab initio} theory development has substantiated two notable points. First, light-matter coupling is typically collective, and increases with the number of charged particles. Second, while the cavity and coupling may appear macroscopic, the collective coupling can still modify the microscopic or local properties, such as the polarizability~\cite{sidler_2020,schnappinger_2023}, due to the formation of an ensemble correlated wave function.

\subsection{Effective models}
\label{subsec:effectivemodels} 

Despite progress with the first-principles approach, it is often advantageous to distill the relevant physics into effective models. This is typically achieved either by coupling together a restricted set of matter and light degrees of freedom, or by reducing the already coupled light-matter system in the PF theory to a simplified low-energy model. Both approaches must treat the many-body degrees of freedom and emergent fluctuations of cavity quantum materials, and go beyond the paradigmatic quantum Rabi model.   

Following the first strategy, established matter models of correlated quantum materials, such as Hubbard or spin Hamiltonians, are coupled to selected photon degrees of freedom. Many proposals extend condensed-matter Hamiltonians by introducing quantized photon modes with ad hoc couplings, making the form of light-matter coupling a central issue. The interaction is typically introduced via minimal coupling $\hat{p}\rightarrow \hat{p}-e\hat{A}(\bm{r},t)$, albeit for effective particles instead of the original bare ones featured in first principles calculations (see Sec.~\hyperref[subsec:firstprinciples]{First principles}). In analogy to the PF Hamiltonian, expansion results in linear ($A$) and quadratic terms ($A^2$), paramagnetic and diamagnetic coupling, respectively. While in many traditional cavity QED situations the $A^2$ term can be neglected (see \hyperref[theorybox]{Theory Overview Box}~(B)), both terms become relevant in the ultrastrong coupling regime~\cite{li2018vacuum} (see \hyperref[theorybox]{Theory Overview Box}~(C)). In this regime, for example, nonzero photon occupations and higher orbitals contribute substantially to the ground state~\cite{lu.shin.ea_2025}. 

The second approach is to start from the light-matter \textit{ab initio} PF Hamiltonian and downfold the system to low-energy models. In the long-wavelength approximation, this downfolding is performed by applying unitary transformations that mix light and matter, such as the Power–Zienau–Woolley (PZW) transformation~\cite{bernardis2018}
or adiabatic decoupling~\cite{Ashida2021}. Multi-center PZW transformations have been developed to dress localized Wannier orbitals with quantized electromagnetic fields, yielding fully quantum Hamiltonians where light and electrons couple through Peierls phases and electric dipole matrix elements (see \hyperref[theorybox]{Theory Overview Box}~(D)).

In both approaches, projecting the light–matter coupling onto a limited set of material and cavity degrees of freedom can produce effective cavity modes with mixed longitudinal and transverse character~\cite{Keeling_2007,Rabl_schuler_vacua_2020} (cf. \hyperref[theorybox]{Theory Overview Box}~(E)), however an open challenge in the field is to maintain equal treatment of these mixed modes. 

State-of-the-art approaches to solving these many-body problems follow a few different routes. First, one can integrate out the photon field, formulating the theory in terms of electronic $G(\bm{r},\bm{r}',t)$ and photon propagators $D(\bm{r},\bm{r}',t)$~\cite{deMelo2016}. The result is a matter-only Hamiltonian that mirrors conventional electronic structure theory, and is well-suited to systems with itinerant charge degrees of freedom. This approach is currently only possible in the weak coupling regime, and handling truly hybrid light-matter modes remains an area of open research. In the limit of gapped charge dynamics, the light–matter coupling can be described as interactions between cavity fluctuations and localized spins. In this regime, one can downfold electronic Hamiltonians to spin–photon models using cavity Schrieffer–Wolff transformations~\cite{sentef.li.ea_2020}. The resulting effective Hamiltonians contain nonlinear spin–photon couplings, with close phenomenological links to nonlinear optical probes such as Raman scattering. However, Schrieffer–Wolff transformations rely on a clear energy scale separation and developing near-resonant downfolding schemes is an intriguing theoretical challenge.

While the most straightforward method to numerically solve the resulting many-body Hamiltonian is exact diagonalization, this method scales poorly and is limited in use in multi-mode systems~\cite{li.eckstein_2020}. Entangling transformations~\cite{ashioda_waveguide_2022} and tensor-network density-matrix renormalization groups ~\cite{passetti_cavity_2023,sanchez-burillo2014} have been implemented to make larger system sizes affordable, but these are limited in their application to bosons. Other advances have been made in ground state solvers like quantum Monte Carlo techniques~\cite{Weber_qmc_spin_boson_2022,aaram_kim_vertex_2023,langheld2024quantumphasediagramsdickeising}, or systematic $1/N$ expansions~\cite{zueco_2025_linear_response}. However, scaling these tools to fluctuating multimode cavities coupled to realistic correlated solids remains an area of open research.

\subsection{Common challenges}

In both effective models and \textit{ab initio} approaches, one typically selects a subset of cavity modes that interact with the electronic or nuclear subsystem. This often neglects the fact that the material itself will reshape the mode functions. For example, a metal imposes a node in the mode profile, effectively making the interacting field a superposition of multiple bare modes. In principle, one therefore should consider the modes self-consistently with the matter. This has been achieved for \textit{ab initio} treatments of molecular systems~\cite{bustamante2025molecular}, but  requires further application to quantum materials. In effective models, downfolding from \textit{ab initio} theory would need to be done self-consistently, identifying the relevant light–matter modes while avoiding double-counting of interactions (see \hyperref[theorybox]{Theory Overview Box}~(E)).

Another area of active theoretical development are approaches that go beyond the long-wavelength approximation, as required for the current generation of cavity quantum materials with large electric field gradients or itinerant electron systems~\cite{enknergraziottofraction}. \textit{Ab initio}, self‑consistent frameworks have been demonstrated for nonchiral molecular systems~\cite{Bonafe2025}, providing a foundation for investigating twisted light, orbital‑angular‑momentum effects, and other cavity-modified properties complex materials.

\onecolumngrid \label{theorybox}
\begin{tcolorbox}[breakable,
  colframe=blue!40!black,
  colback=blue!8,
  coltitle=white,
  title=\Large Theory Overview: Representative Hamiltonians and Couplings$^{\ast}$,
  fonttitle=\bfseries,
  left=10pt,right=10pt,top=10pt,bottom=10pt,
  width=\textwidth]

\textbf{(A) Quantum-optics Rabi model (single emitter, single mode).}
\begin{align}
H_{\mathrm{Rabi}}
&= \hbar\omega_c\, a^\dagger a + \frac{\hbar\omega_q}{2}\,\sigma_z
+ \hbar g\,(\sigma_+ + \sigma_-)(a + a^\dagger).
\end{align}
$a$ ($a^\dagger$): cavity-photon annihilation (creation);  
$\sigma_z, \sigma_\pm$: Pauli operators for a two-level system;  
$\omega_c$: effective cavity frequency; $\omega_q$: qubit/emitter frequency; $g$: light–matter coupling.  
\textbf{Coupling regimes:} weak $g<\{\kappa,\gamma\}$; strong $g>\{\kappa,\gamma\}$ ($\kappa,\gamma$: cavity and qubit/emitter loss rates); ultrastrong $0.1 \lesssim g/\omega_{c,q} < 1$ (counter-rotating terms and $A^2$ required); deep-strong $g/\omega_{c,q}>1$. The model can be trivially extended to many modes and multiple non-overlapping emitters.  

\vspace{4pt}\hrule\vspace{4pt}

\textbf{(B) Multimode cavity coupled to collective bosons (e.g., IR active phonons/excitons).}
\begin{align}
H_{\mathrm{bos}}
&= H_{\rm cav}
  + \sum_{\mathbf{q}}\hbar\Omega_{\mathbf{q}}\, b^\dagger_{\mathbf{q}} b_{\mathbf{q}}
  + \sum_{\mu,\mathbf{q}} \hbar g_{\mu\mathbf{q}}
    \,(a_{\mu} + a^\dagger_{\mu})(b_{\mathbf{q}} + b^\dagger_{\mathbf{q}}).
\end{align}
$b_{\mathbf{q}}$ ($b^\dagger_{\mathbf{q}}$): bosonic excitations (IR-active phonons, excitons) with momentum $\mathbf{q}$.  
Multimode or near-field cavities $\Rightarrow$ dense $\{\omega_\mu\}$ and broad density of states. Couplings $g_{\mu\mathbf{q}}$ depend on overlap between photon mode functions $\mathbf{u}_\mu(\mathbf{r})$ and material polarization $\mathbf{P}_{\mathbf{q}}(\mathbf{r})$ ($H_{\mathrm{cav}}$ as in (E)).  

\vspace{4pt}\hrule\vspace{4pt}

\textbf{(C) Low-energy itinerant electrons in a cavity (minimal coupling).}
\begin{align}
H &= H_{\mathrm{el}} + H_{\mathrm{cav}} + H_{\mathrm{para}} + H_{\mathrm{diam}} + H_{\phi}, \\[-4pt]
H_{\mathrm{para}} &= \sum_{\mathbf{k}} \mathbf{j}_{\mathbf{k}}\!\cdot\!\mathbf{A}_{\rm cav}, \quad
H_{\mathrm{diam}} \propto \mathbf{A}^2_{\rm cav}, \quad
H_{\phi} = \sum_{\mathbf{q}} \rho_{-\mathbf{q}}\,\phi^{\mathbf{q}}_{\rm cav}.
\end{align}
$\mathbf{j}_{\mathbf{k}}$: electronic paramagnetic current operator; $\rho_{\mathbf{q}}$: Fourier component of electron density.  
Cavity-photon exchange mediates effective interactions at the Fermi surface.  
($\mathbf{A}_{\rm cav}, \phi_{\rm cav}, H_{\mathrm{cav}}$ as in (E)).  

\vspace{4pt}\hrule\vspace{4pt}

\textbf{(D) Strongly correlated lattice: Hubbard model with quantized Peierls substitution.}
\begin{align}
H_{\mathrm{Hub}}[\mathbf{A}_{\rm cav}]
&= -\!\!\sum_{\langle ij\rangle,\sigma}\! t_{ij}\,
e^{\,\frac{i e}{\hbar}\int_{\mathbf{R}_i}^{\mathbf{R}_j}\!\mathbf{A_{\rm cav}}(\mathbf{r})\cdot d\boldsymbol{\ell}}\,
c^\dagger_{i\sigma} c_{j\sigma}
+ U \sum_i n_{i\uparrow} n_{i\downarrow}
- \mu \sum_{i,\sigma} n_{i\sigma}
+ H_{\mathrm{cav}}.
\end{align}
Cavity-mode operators enter through the Peierls phases (and through dynamical scalar potential, not shown), enabling cavity-mediated
renormalization of $t$, $U$, and model correlated phases (e.g.\ superconductivity, magnetism). Starting point for effective spin–photon Hamiltonians.  
For multi-orbital models, downfolding (e.g.\ via multi-center PZW) generates couplings between localized Wannier orbitals and the cavity field.  
($\mathbf{A}_{\rm cav}, H_{\rm cav}$ as in (E)).  
\vspace{4pt}\hrule\vspace{4pt}

\textbf{(E) Effective cavity fields.}
\begin{align}
H_{\mathrm{cav}} &= \sum_{\mu}\hbar\omega_\mu \Big(a_\mu^\dagger a_\mu + \tfrac{1}{2}\Big), \quad
\mathbf{A}_{\rm cav}(\mathbf{r}) = \sum_{\mu}\sqrt{\tfrac{\hbar}{2\epsilon_0 \omega_\mu}}\,\mathbf{u}_\mu(\mathbf{r})\,(a_\mu + a_\mu^\dagger), \quad
\phi_{\rm cav}(\mathbf{r}) = \sum_{\mu}\sqrt{\tfrac{\hbar}{2\epsilon_0 \omega_\mu}}\,v_\mu(\mathbf{r})\,(a_\mu + a_\mu^\dagger).
\end{align}
If we approximate the part of the Pauli-Fierz Hamiltonian (as defined in (F)) that describes the cavity structure, we can introduce effective cavity (hybrid light-matter) modes $\mu$. These modes are not automatically purely transverse or longitudinal, and one sometimes subsumes the effect of the unmodified free-space modes into the \emph{observable} masses of the particles. 

\vspace{4pt}\hrule\vspace{4pt}

\textbf{(F) \textit{Ab initio} starting point: Pauli-Fierz Hamiltonian (Coulomb gauge).}
\begin{equation}
\begin{aligned}
H_{\mathrm{PF}}
&= \sum_i \left\{\frac{\big[\mathbf{p}_i + |e|\,\mathbf{A}_{\perp}(\mathbf{r}_i)\big]^2}{2 m_e}+\frac{|e|\hbar}{2m_{e}}\boldsymbol{\sigma}_{i}\cdot\mathbf{B}(\mathbf{r}_{i})\right\}
+ V_{\mathrm{Coul}}(\{\mathbf{r}_i\})
+ \sum_{\lambda =1}^{2} \int \hbar \omega_{k} a^{\dagger}(\mathbf{k},\lambda)a^{}(\mathbf{k},\lambda) \mathrm{d} \mathbf{k} \\
& + \sum_I \left\{\frac{\big[\mathbf{P}_I - Z_{I}|e|\,\mathbf{A}_{\perp}(\mathbf{R}_I)\big]^2}{2 M_I}-\frac{Z_{I}|e|\hbar}{2M_{I}}\boldsymbol{S}_{I}\cdot\mathbf{B}(\mathbf{R}_{i})\right\} + V_{\rm{Coul}}(\left\{\mathbf{R}_{I}\right\}) + V_{\rm{Coul}}(\left\{\mathbf{r}_{i},\mathbf{R}_{I}\right\}).
\end{aligned}
\end{equation}
$\mathbf{p}_i$ ($\mathbf{P}_I$): electronic (nuclear) momentum; $e$: the \emph{observable} magnitude of electron charge; $Z_{I}|e|$: nuclear charge; $m_{e}$ ($M_{I}$): \emph{bare} electron (nuclear) mass; 
\begin{align}
\mathbf{A}_{\perp}(\mathbf{r}) = \sum_{\lambda =1}^{2}\int \sqrt{\tfrac{\hbar}{2\epsilon_0 \omega_k}}\left( a(\mathbf{k},\lambda) \mathbf{u}_{\mathbf{k},\lambda}(\mathbf{r}) + a^{\dagger}(\mathbf{k} ,\lambda) \mathbf{u}^{*}_{\mathbf{k},\lambda}(\mathbf{r})\right) \mathrm{d} \mathbf{k}:
\end{align}
transverse vector potential with $\mathbf{u}_{\mathbf{k},\lambda}(\mathbf{r})= \boldsymbol{\epsilon}(\mathbf{k},\lambda)\exp(\mathrm{i} \mathbf{k} \cdot \vec{r})/(2\pi)^{3/2}$, $\boldsymbol{\epsilon}(\mathbf{k},\lambda)$ the two transverse polarization vectors and $\omega_k = c|\mathbf{k}|$;  $\mathbf{B} = \mathbf{\nabla}\times\mathbf{A}_{\perp}$: quantized magnetic field; $\boldsymbol{\sigma}_{i}$ ($\boldsymbol{S}_{I}$): electronic (nuclear) spin matrices; $V_{\mathrm{Coul}}$: longitudinal Coulomb interaction. Provides rigorous QED foundation, including the full transverse (radiative) and longitudinal (Coulomb) fields coupled to the \emph{bare} electrons and nuclei. Basis for first-principle methods such as QEDFT or QED coupled-cluster theory and parent Hamiltonian for downfolding to effective low-energy models (for effective cavity modes as in (E) and effective light-matter Hamiltonians as in (A)-(D)).    
\vspace{6pt}\hrule\vspace{3pt}
\centering
{\footnotesize $^{\ast}$This overview provides a rough guide to representative models and is not exhaustive.}

\end{tcolorbox}
\twocolumngrid

\section{Prospects for cavity control}\label{sec:outlook}
\subsection{Equilibrium control of established phases}\label{sec:outlookphases}
While cavity engineering of a wide range of quantum phases has been proposed, we review promising research directions in fluctuation engineering of superconductivity, magnetism, and topology.

The recent experiment on $\kappa$-ET interfaced with hBN~\cite{keren2025cavity} showcases one example of how cavity embedding can affect superconductivity. Different pairing mechanisms are expected to require different routes for cavity control. For phonon-mediated superconductors, coupling to cavity photons can hybridize lattice vibrations into phonon–polaritons, thereby reshaping the spectrum of bosons that mediate pairing. For spin-fluctuation-driven superconductors, cavities offer the possibility of influencing magnetic excitations indirectly, for example via cavity-modified exchange pathways or by tuning low-energy collective modes through anisotropic fields. In intrinsically strong-coupling superconductors, where pairing arises from correlations in a non-Fermi-liquid background, direct electron–photon coupling can alter the kinetic energy balance and effective bandwidth, with consequences for the stability of superconducting order. 


Quantum magnets are another natural class of materials amenable to cavity fluctuation engineering. Theoretical works suggest that cavity fluctuations can drive $\alpha$-RuCl$_3$ from a zigzag antiferromagnet to a ferromagnet, induce finite magnetization in FePS$_3$, and tune the spiral vector of NiI$_2$, thereby altering its multiferroic properties~\cite{vinas_bostrom_controlling_2023,masuki.ashida_2023}. These effects originate from cavity-induced renormalization of exchange interactions mediated by virtual electronic processes and phonon–polariton couplings. Spiral magnets such as NiI$_2$~\cite{gao2024giant} are particularly sensitive platforms, since small cavity-mediated changes in exchange couplings translate directly into measurable shifts of spiral momentum. Harnessing such mechanisms could establish THz cavities as a powerful and energy-efficient tool for reconfiguring magnetic orders and engineering exotic quantum states in strongly correlated materials.

Another compelling research direction is to use cavity engineering to manipulate the topological properties of quantum materials, such as breaking time-reversal symmetry (TRS) by embedding a material in a chiral cavity~\cite{wang2019cavity,hubener2021engineering}. Theoretical proposals suggest that such cavities could induce a variety of topological phases, including vacuum-fluctuation–driven Chern insulators in graphene~\cite{wang2019cavity}, flattened bands in twisted bilayer graphene~\cite{jiang2024engineering}, or cavity-induced fractional Chern insulators~\cite{nguyen2023electron}. To break TRS via vacuum fluctuations, right- and left-circularly polarized modes must be non-degenerate, forming a so-called Faraday mode~\cite{hubener2021engineering}. This configuration is distinct from helical cavities—such as metamaterial designs—that selectively reflect a single-handedness of light without intrinsically breaking TRS. TRS breaking has recently been demonstrated in various experimental platforms~\cite{aupiais2024chiral-908,tay2025terahertz,andberger2024terahertz,suarez-forero_chiral_2024}, but an open experimental challenge is realizing cavities whose chirality is intrinsic without the need of an external magnetic fields to determine the handedness.

\subsection{Non-equilibrium fluctuation-driven regimes}

A second frontier of cavity quantum materials lies in non-equilibrium control. Despite having close ties since its inception with Floquet engineering, many questions still remain regarding the distinction between coherent and incoherent few-photon drives, as well as quantum versus thermal bath engineering. Cavities provide a platform to merge with Floquet approaches and to stabilize genuinely non-equilibrium phases, offering novel ways to overcome limitations of classical drives and to harness or engineer dissipation to generate persistent metastability. 


For instance, one natural angle for cavity engineering is to replace classical light with quantum light while preserving the toolbox of Floquet engineering. In principle, there should be a quantum-to-classical crossover between cavity engineering -- using enhanced light-matter coupling with virtual ``photons'' -- and Floquet engineering -- using free-space coupling to coherent real photons. Early theory works on the quantum-to-classical crossover of Floquet engineering in the Hubbard model~\cite{sentef.li.ea_2020} and in exactly solvable chains~\cite{eckhardt2022quantum} showed that typical limitations of Floquet engineering, like heating, decoherence, and transience, could be mitigated in a cavity. Remarkably, even incoherent single-photon states can reproduce Floquet-like band modifications if the effective single-particle coupling strength is large enough~\cite{sentef.li.ea_2020}, such that nearly all classical ``Floquet proposals'' can be recast as ``Floquet-cavity proposals'' and open opportunities distinct from cavity or Floquet scenario alone. 

One challenge is to reach regimes where such Floquet-like phenomena become observable in real cavity–material platforms. Cavity field enhancements or reduced Brillouin zones~\cite{de_la_torre_colloquium_2021} could facilitate attainment of the quantum Floquet regime. Experiments using plasmonic\cite{reutzel_nonlinear_2019} and photonic~\cite{zhou_cavity_2024} cavities have enabled Floquet engineering at fluences comparable with continuous-wave sources. Combining near-field circuitry~\cite{mciver2020light} with near-field cavities and continuous wave-drives~\cite{yoshioka2020onchip} offers another promising direction for cavity-Floquet engineering.


An idea distinct from the few-photon limit of cavity-Floquet engineering is to use incoherent cavity drives to generate and stabilize non-equilibrium states of matter. In this picture, the cavity controls radiative exchange between a material degree of freedom and an external photon bath~\cite{pannir2025blackbody,fassioli_fausti_2024}, such that different degrees of freedom remain out of equilibrium with each other. This provides a route to achieve steady states that evade equilibrium constraints such as the fluctuation–dissipation relation. 
Key design parameters thought to be relevant for these fluctuation-driven states include:  
(i) strong single-dipole or collective light–matter coupling, which shapes the hybrid spectrum relative to the bath’s spectral density;  
(ii) cavities engineered as open and dissipative systems, so that the external bath can effectively drive the material through radiative channels;  
(iii) careful balance between cavity-mediated coupling of selected material modes to the photon bath and competing intrinsic relaxation channels within the material. With these designs, dissipative control could play as central as coherent control in determining material properties.

\subsection{Platforms and diagnostics}
\label{subsec-LMCoupling}

A final important frontier for cavity quantum materials is the development of new platforms for investigating and leveraging cavity engineered states. 

One complexity of sub-wavelength cavities is the limited in situ tuning control on the influence of the cavity itself, in contrast to Fabry-Pérot cavities whose cavity mirrors can be easily removed. While mechanical control has been used to impose and remove a split ring resonator cavity~\cite{enknergraziottofraction}, gate-tunable 2D materials could facilitate mapping the entire cavity-modified phase diagram. The enlarged lattice constants---or, equivalently, reduced Brillouin zones---are expected to enhance light–matter coupling, akin to increasing the effective Floquet parameter in periodically driven systems~\cite{de_la_torre_colloquium_2021}. Cavity embedding could be especially effective in reshaping correlated phases of moiré materials. Recent experiments on monolayer graphene illustrate the salience of these ideas~\cite{kipp2024cavity}. In a graphene–graphite heterostructure, the 2D plasma polariton modes of thin graphite were shown to act as a near-field cavity mode for graphene, pushing the system into the ultrastrong coupling regime. The ubiquity of graphite gates in the vdW community raises a provocative question: could dynamical hybridization between collective modes of different layers already be influencing the rich ground-state physics observed across the broader family of van der Waals heterostructures? If so, cavity physics may already be present in disguise, awaiting more deliberate exploitation.  

A second area of future work is to develop quantum optical techniques relevant for quantum materials. Probes that move beyond mean-field observables, such as measuring photon correlations, characterizing squeezing, and sensing or generating single photons could provide a route to diagnose quantum phases of matter. Theory suggests that correlation functions of scattered photons can directly probe spin, charge, and topological orders in an electronic system~\cite{nambiar2024diagnosing}, and that signatures of strongly correlated phases may be encoded in non-classical photon statistics~\cite{grunwald2024cavity,Lysne2023,bradley_quantum_2024}. It remains an experimental challenge to probe these quantities in the relevant THz frequency range~\cite{Entanglement2024} . Recent progress has been made in using electro-optic sampling to directly measure the ground-state electric field variance and correlation ~\cite{riek2015direct, benea2019electric,benea2025electro,spencer2025electrooptic}. The enhanced light-matter coupling could be leveraged to advance these techniques. Realizing and harnessing true quantum resources in the THz regime—through the detection, manipulation, and entanglement of individual THz photons—sets a transformative direction for the field, bridging quantum optics and quantum materials research.

\section{Conclusion}

The field of cavity quantum materials has moved rapidly from conceptual proposals to experimental demonstrations where cavities measurably alter collective electronic phases. A unifying theme is the use of cavities as platforms to structure fluctuations that govern correlated matter -- selectively enhancing, suppressing, or redistributing them. The design space is vast, spanning subwavelength confinement, anisotropy, multi-modal polaritons, polarization engineering, and bath driving, each offering new control knobs over strongly interacting degrees of freedom.

Looking forward, progress will hinge on closing the gap between idealized theoretical models and experimental realities, developing \textit{ab initio} and model many-body tools that can predict cavity-modified phase diagrams, and systematically exploring the role of dissipation, disorder, and multimode coupling. The most exciting opportunities may lie where fluctuations are already dominant—unconventional superconductors, quantum magnets, moiré heterostructures, or topological phases—where cavity embedding could address unsettled questions and stabilize elusive states. Developing new observables to probe cavity quantum materials will be critical to providing mechanistic insight and understanding of the quantum phases themselves---and even deliver new techniques for low-energy quantum technologies.  In this way, cavity quantum materials offer not only a platform for fluctuation engineering, but also a broader paradigm in which the electromagnetic environment becomes a co-designer of quantum matter.

\section{Acknowledgments}
This research was supported in part by grant NSF PHY-2309135 to the Kavli Institute for Theoretical Physics (KITP). We specifically acknowledge the kind hospitality of KITP during the program ``Quantum Optics of Correlated Electron Systems'' in Jan/Feb 2025. MAS was funded by the European Union (ERC, CAVMAT, project no. 101124492). DMK acknowledges funding by the
Deutsche Forschungsgemeinschaft (DFG, German Research Foundation) - 508440990.  JWM is supported by the U.S. Department of Energy, Office of Science, Basic Energy Sciences, under Early Career Award DE-SC0024334. MHM and MH acknowledge support from the Alexander von Humboldt Foundation. HMB acknowledges financial support from the European Union under the Marie Sklodowska-Curie Grant Agreement no. 101062921 (Twist-TOC). LG and JF acknowledge funding from the Swiss National Science Foundation (SNF) (Grant number 10000397). JF acknowledges funding from the European Union under the ERC Advanced Grant ‘‘COLLECTIVE''. MH and AG acknowledge DARPA HR00112530313 and ARO W911NF2510066 grants. DF and AM acknowledge support from the Gordon and Betty Moore foundation through the grant CENTQC (no. GBMF12213). AR was supported by the European Research Council (ERC-2024-SyG- 101167294 ; UnMySt), the Cluster of Excellence Advanced Imaging of Matter (AIM). MC acknowledges support from the U.S. Department of Energy, Office of Basic Energy Sciences, under Award No. DE-SC0024494. DMK, HMB, MAS, DF and ME acknowledge funding by the Deutsche Forschungsgemeinschaft (DFG, German Research Foundation)- 531215165 (Research Unit ‘OPTIMAL’)). DNB is supported by DOE-BES grant DE-SC0018426, the Moore Foundation EPIQS award GBMF9455 and ARO grant W911NF2510062.  We acknowledge support from the Max Planck-New York City Center for Non-Equilibrium Quantum Phenomena. The Flatiron Institute is a division of the Simons Foundation. Views and opinions expressed are however those of the author(s) only and do not necessarily reflect those of the European Union or the European Research Council. Neither the European Union nor the European Research Council can be held responsible for them.

\bibliography{references}

@article{andberger2024terahertz, 
  year     = {2024}, 
  title    = {Terahertz chiral subwavelength cavities breaking time-reversal symmetry via ultrastrong light-matter interaction}, 
  author   = {Andberger, Johan and Graziotto, Lorenzo and Sacchi, Luca and Beck, Mattias and Scalari, Giacomo and Faist, Jérôme}, 
  journal  = {Physical Review B}, 
  issn     = {2469-9950}, 
  doi      = {10.1103/physrevb.109.l161302}, 
  pages    = {L161302}, 
  number   = {16}, 
  volume   = {109}
}

@article{gogna2020self,
  title={Self-hybridized, polarized polaritons in {ReS}$_2$ crystals},
  author={Gogna, Rahul and Zhang, Long and Deng, Hui},
  journal={ACS Photonics},
  volume={7},
  number={12},
  pages={3328--3332},
  year={2020},
  publisher={ACS Publications},
    doi = {10.1021/acsphotonics.0c01537}
}

@article{scalari2012ultrastrong,
  title={Ultrastrong coupling of the cyclotron transition of a {2D} electron gas to a {THz} metamaterial},
  author={Scalari, Giacomo and Maissen, Curdin and Tur{\v{c}}inkov{\'a}, Dana and Hagenm{\"u}ller, David and De Liberato, Simone and Ciuti, Cristiano and Reichl, Christian and Schuh, Dieter and Wegscheider, Werner and Beck, Mattias and others},
  journal={Science},
  volume={335},
  number={6074},
  pages={1323--1326},
  year={2012},
  publisher={American Association for the Advancement of Science},
    doi ={10.1126/science.1216022}
}

@article{tay2025terahertz, 
  year     = {2025}, 
  title    = {Terahertz chiral photonic-crystal cavities for {Dirac} gap engineering in graphene}, 
  author   = {Tay, Fuyang and Sanders, Stephen and Baydin, Andrey and Song, Zhigang and Welakuh, Davis M. and Alabastri, Alessandro and Rokaj, Vasil and Dag, Ceren B. and Kono, Junichiro}, 
  journal  = {Nature Communications}, 
  doi      = {10.1038/s41467-025-60335-x}, 
  pmid     = {40481009}, 
  pmcid    = {{PMC}12144285}, 
  eprint   = {2410.21171}, 
  pages    = {5270}, 
  number   = {1}, 
  volume   = {16}
}

@article{Lysne2023,
  title = {Quantum Optics Measurement Scheme for Quantum Geometry and Topological Invariants},
  author = {Lysne, Markus and Sch\"uler, Michael and Werner, Philipp},
  journal = {Phys. Rev. Lett.},
  volume = {131},
  issue = {15},
  pages = {156901},
  numpages = {7},
  year = {2023},
  month = {Oct},
  publisher = {APS},
  doi = {10.1103/PhysRevLett.131.156901},
  url = {https://link.aps.org/doi/10.1103/PhysRevLett.131.156901}
}

@article{nambiar2024diagnosing,
  title = {Diagnosing Electronic Phases of Matter Using Photonic Correlation Functions},
  author = {Nambiar, Gautam and Grankin, Andrey and Hafezi, Mohammad},
  journal = {Phys. Rev. X},
  volume = {15},
  issue = {4},
  pages = {041020},
  numpages = {51},
  year = {2025},
  month = {Nov},
  publisher = {American Physical Society},
  doi = {10.1103/67zs-hqf3},
  url = {https://link.aps.org/doi/10.1103/67zs-hqf3}
}

@article{sternbach2020femtosecond,
  title={Femtosecond exciton dynamics in {WSe}$_2$ optical waveguides},
  author={Sternbach, Aaron J and Latini, Simone and Chae, Sanghoon and H{\"u}bener, Hannes and De Giovannini, Umberto and Shao, Yinming and Xiong, Lin and Sun, Zhiyuan and Shi, Norman and Kissin, Peter and others},
  journal={Nature communications},
  volume={11},
  number={1},
  pages={3567},
  year={2020},
  publisher={Nature Publishing Group UK London}, 
    doi = {10.1038/s41467-020-17335-w}
}

@article{ebbesen2016hybrid,
  title={Hybrid light--matter states in a molecular and material science perspective},
  author={Ebbesen, Thomas W},
  journal={Accounts of chemical research},
  volume={49},
  number={11},
  pages={2403--2412},
  year={2016},
  publisher={ACS Publications},
    doi = {10.1021/acs.accounts.6b00295}
}

@article{spencer2025electrooptic, 
  year     = {2025}, 
  title    = {Electro-optic cavities for in-situ measurement of cavity fields}, 
  author   = {Spencer, Michael S. and Urban, Joanna M. and Frenzel, Maximilian and Mueller, Niclas S. and Minakova, Olga and Wolf, Martin and Paarmann, Alexander and Maehrlein, Sebastian F.}, 
  journal  = {Light: Science \& Applications}, 
  issn     = {2095-5545}, 
  doi      = {10.1038/s41377-024-01685-x}, 
  pmid     = {39910060}, 
  pmcid    = {{PMC}11799363}, 
  pages    = {69}, 
  number   = {1}, 
  volume   = {14}
}

@article{session2025optical,
  title={Optical pumping of electronic quantum {Hall} states with vortex light},
  author={Session, Deric and Jalali Mehrabad, Mahmoud and Paithankar, Nikil and Grass, Tobias and Eckhardt, Christian J and Cao, Bin and Gustavo Su{\'a}rez Forero, Daniel and Li, Kevin and Alam, Mohammad S and Watanabe, Kenji and others},
  journal={Nature Photonics},
  volume={19},
  number={2},
  pages={156--161},
  year={2025},
  publisher={Nature Publishing Group UK London},
    doi ={10.1038/s41566-024-01565-1}
}

@article{Ashida2021,
  title = {Cavity Quantum Electrodynamics at Arbitrary Light-Matter Coupling Strengths},
  author = {Ashida, Yuto and Imamoglu, Atac and Demler, Eugene},
  journal = {Phys. Rev. Lett.},
  volume = {126},
  issue = {15},
  pages = {153603},
  numpages = {8},
  year = {2021},
  month = {Apr},
  publisher = {American Physical Society},
  doi = {10.1103/PhysRevLett.126.153603},
  url = {https://link.aps.org/doi/10.1103/PhysRevLett.126.153603}
}

@article{aaram_kim_vertex_2023,
  title = {Vertex-Based Diagrammatic Treatment of Light-Matter-Coupled Systems},
  author = {Kim, Aaram J. and Lenk, Katharina and Li, Jiajun and Werner, Philipp and Eckstein, Martin},
  journal = {Phys. Rev. Lett.},
  volume = {130},
  issue = {3},
  pages = {036901},
  numpages = {6},
  year = {2023},
  month = {Jan},
  publisher = {American Physical Society},
  doi = {10.1103/PhysRevLett.130.036901},
  url = {https://link.aps.org/doi/10.1103/PhysRevLett.130.036901}
}

@article{jarc2022tunable,
  title={Tunable cryogenic terahertz cavity for strong light--matter coupling in complex materials},
  author={Jarc, Giacomo and Mathengattil, Shahla Yasmin and Giusti, Francesca and Barnaba, Maurizio and Singh, Abhishek and Montanaro, Angela and Glerean, Filippo and Rigoni, Enrico Maria and Zilio, Simone Dal and Winnerl, Stephan and others},
  journal={Review of Scientific Instruments},
  volume={93},
  number={3},
  year={2022},
    doi = {10.1063/5.0080045},
  publisher={AIP Publishing}
}

@article{ashioda_waveguide_2022,
  title = {Nonperturbative waveguide quantum electrodynamics},
  author = {Ashida, Yuto and Yokota, Takeru and  {\.I}mamo{\u{g}}lu, Ata{\c{c}} and Demler, Eugene},
  journal = {Phys. Rev. Res.},
  volume = {4},
  issue = {2},
  pages = {023194},
  numpages = {26},
  year = {2022},
  month = {Jun},
  publisher = {American Physical Society},
  doi = {10.1103/PhysRevResearch.4.023194},

}

@article{Keeling_2007,
doi = {10.1088/0953-8984/19/29/295213},
url = {https://dx.doi.org/10.1088/0953-8984/19/29/295213},
year = {2007},
month = {jun},
publisher = {},
volume = {19},
number = {29},
pages = {295213},
author = {Keeling, Jonathan},
title = {Coulomb interactions, gauge invariance, and phase transitions of the {Dicke} model},
journal = {Journal of Physics: Condensed Matter}
}

@Article{Rabl_schuler_vacua_2020,
	title={{The vacua of dipolar cavity quantum electrodynamics}},
	author={Michael Schuler and Daniele De Bernardis and Andreas M. Läuchli and Peter Rabl},
	journal={SciPost Phys.},
	volume={9},
	pages={066},
	year={2020},
	publisher={SciPost},
	doi={10.21468/SciPostPhys.9.5.066},
	url={https://scipost.org/10.21468/SciPostPhys.9.5.066},
}

@article{Weber_qmc_spin_boson_2022,
  title = {{{Quantum Monte Carlo}} simulation of spin-boson models using wormhole updates},
  author = {Weber, Manuel},
  journal = {Phys. Rev. B},
  volume = {105},
  issue = {16},
  pages = {165129},
  numpages = {14},
  year = {2022},
  month = {Apr},
  publisher = {American Physical Society},
  doi = {10.1103/PhysRevB.105.165129},
  url = {https://link.aps.org/doi/10.1103/PhysRevB.105.165129}
}

@article{bernardis2018,
  title = {Cavity quantum electrodynamics in the nonperturbative regime},
  author = {De Bernardis, Daniele and Jaako, Tuomas and Rabl, Peter},
  journal = {Phys. Rev. A},
  volume = {97},
  issue = {4},
  pages = {043820},
  numpages = {18},
  year = {2018},
  month = {Apr},
  publisher = {American Physical Society},
  doi = {10.1103/PhysRevA.97.043820},
  url = {https://link.aps.org/doi/10.1103/PhysRevA.97.043820}
}

@article{langheld2024quantumphasediagramsdickeising,
  title = {Quantum phase diagrams of {Dicke-Ising} models by a wormhole algorithm},
  author = {Langheld, Anja and H\"ormann, Max and Schmidt, Kai Phillip},
  journal = {Phys. Rev. B},
  volume = {112},
  issue = {16},
  pages = {L161123},
  numpages = {7},
  year = {2025},
  month = {Oct},
  publisher = {APS},
  doi = {10.1103/lcvj-ksct},
  url = {https://link.aps.org/doi/10.1103/lcvj-ksct}
}

@article{zueco_2025_linear_response,
  title = {Linear response theory for cavity {QED} materials at arbitrary light-matter coupling strengths},
  author = {Rom\'an-Roche, Juan and G\'omez-Le\'on, \'Alvaro and Luis, Fernando and Zueco, David},
  journal = {Phys. Rev. B},
  volume = {111},
  issue = {3},
  pages = {035156},
  numpages = {21},
  year = {2025},
  month = {Jan},
  publisher = {American Physical Society},
  doi = {10.1103/PhysRevB.111.035156},
  url = {https://link.aps.org/doi/10.1103/PhysRevB.111.035156}
}

@article{sanchez-burillo2014,
  title = {Scattering in the Ultrastrong Regime: Nonlinear Optics with One Photon},
  author = {Sanchez-Burillo, E. and Zueco, D. and Garcia-Ripoll, J. J. and Martin-Moreno, L.},
  journal = {Phys. Rev. Lett.},
  volume = {113},
  issue = {26},
  pages = {263604},
  numpages = {5},
  year = {2014},
  month = {Dec},
  publisher = {American Physical Society},
  doi = {10.1103/PhysRevLett.113.263604},
  url = {https://link.aps.org/doi/10.1103/PhysRevLett.113.263604}
}

@article{Bonafe2025,
  title = {Full minimal coupling {Maxwell-TDDFT:} An ab initio framework for light-matter interaction beyond the dipole approximation},
  author = {Bonaf\'e, Franco P. and Albar, Esra Ilke and Ohlmann, Sebastian T. and Kosheleva, Valeriia P. and Bustamante, Carlos M. and Troisi, Francesco and Rubio, Angel and Appel, Heiko},
  journal = {Phys. Rev. B},
  volume = {111},
  issue = {8},
  pages = {085114},
  numpages = {14},
  year = {2025},
  month = {Feb},
  publisher = {American Physical Society},
  doi = {10.1103/PhysRevB.111.085114},
  url = {https://link.aps.org/doi/10.1103/PhysRevB.111.085114}
}

@article{svendsen_2023,
  title={Effective equilibrium theory of quantum light-matter interaction in cavities for extended systems and the long wavelength approximation},
  author={Svendsen, Mark Kamper and Ruggenthaler, Michael and H{\"u}bener, Hannes and Sch{\"a}fer, Christian and Eckstein, Martin and Rubio, Angel and Latini, Simone},
  journal={Communications Physics},
  volume={8},
  number={1},
  pages={425},
  year={2025},
  publisher={Nature Publishing Group UK London},
    doi ={10.1038/s42005-025-02365-x}
}

@article{dai2014tunable,
  title={Tunable phonon polaritons in atomically thin van der Waals crystals of boron nitride},
  author={Dai, Siyuan and Fei, Z and Ma, Q and Rodin, AS and Wagner, M and McLeod, AS and Liu, MK and Gannett, W and Regan, W and Watanabe, K and others},
  journal={Science},
  volume={343},
  number={6175},
  pages={1125--1129},
  year={2014},
  publisher={American Association for the Advancement of Science},
    doi = {10.1126/science.1246833}
}

@article{aupiais2024chiral-908, 
  year     = {2024}, 
  title    = {Chiral {TeraHertz} Surface Plasmonics}, 
  author   = {Aupiais, Ian and Grasset, Romain and Daineka, Dmitri and Briatico, Javier and Perfetti, Luca and Hugonin, Jean-Paul and Greffet, Jean-Jacques and Laplace, Yannis}, 
  journal  = {{ACS} Photonics}, 
  issn     = {2330-4022}, 
  doi      = {10.1021/acsphotonics.4c01076}, 
  eprint   = {2404.15270}, 
  pages    = {4184--4192}, 
  number   = {10}, 
  volume   = {11}
}

@article{keren2025cavity,
  title={Cavity-altered superconductivity},
  author={Keren, Itai and Webb, Tatiana A and Zhang, Shuai and Xu, Jikai and Sun, Dihao and Kim, Brian SY and Shin, Dongbin and Zhang, Songtian S and Zhang, Junhe and Pereira, Giancarlo and others},
  journal={Nature},
  volume={650},
  number={8103},
  pages={864--868},
  year={2026},
  doi = {10.1038/s41586-025-10062-6},
  publisher={Nature Publishing Group}
}

@article{dirnberger2023magneto,
  title={Magneto-optics in a van der {{W}}aals magnet tuned by self-hybridized polaritons},
  author={Dirnberger, Florian and Quan, Jiamin and Bushati, Rezlind and Diederich, Geoffrey M and Florian, Matthias and Klein, Julian and Mosina, Kseniia and Sofer, Zdenek and Xu, Xiaodong and Kamra, Akashdeep and others},
  journal={Nature},
  volume={620},
  number={7974},
  pages={533--537},
  year={2023},
  publisher={Nature Publishing Group UK London},
    doi = {10.1038/s41586-023-06275-2}
}

@article{eckhardt2024surface,
  title = {Surface-Mediated Ultrastrong Cavity Coupling of Two-Dimensional Itinerant Electrons},
  author = {Eckhardt, Christian J. and Grankin, Andrey and Kennes, Dante M. and Ruggenthaler, Michael and Rubio, Angel and Sentef, Michael A. and Hafezi, Mohammad and Michael, Marios H.},
  journal = {Phys. Rev. Lett.},
  volume = {135},
  issue = {15},
  pages = {156902},
  numpages = {7},
  year = {2025},
  month = {Oct},
  publisher = {American Physical Society},
  doi = {10.1103/2fw2-lbhy},
  url = {https://link.aps.org/doi/10.1103/2fw2-lbhy}
}

@article{kipp2024cavity,
  title={Cavity electrodynamics of van der {{W}}aals heterostructures},
  author={Kipp, Gunda and Bretscher, Hope M and Schulte, Benedikt and Herrmann, Dorothee and Kusyak, Kateryna and Day, Matthew W and Kesavan, Sivasruthi and Matsuyama, Toru and Li, Xinyu and Langner, Sara Maria and others},
  journal={Nature Physics},
  year={2025},
      volume={15},
  number={19},
  pages={1926––1933},
    doi  = {10.1038/s41567-025-03064-8}, 
    Publisher = {Nature Publishing Group UK London}
    
}

@article{sidler_2020,
  title={Polaritonic chemistry: Collective strong coupling implies strong local modification of chemical properties},
  author={Sidler, Dominik and Sch{\"a}fer, Christian and Ruggenthaler, Michael and Rubio, Angel},
  journal={{The Journal of Physical Chemistry Letters}},
  volume={12},
  number={1},
  pages={508--516},
  year={2020},
  publisher={ACS Publications},
    doi = {10.1021/acs.jpclett.0c03436}
}

@article{schnappinger_2023,
  title={Cavity Born--Oppenheimer Hartree--Fock ansatz: Light--matter properties of strongly coupled molecular ensembles},
  author={Schnappinger, Thomas and Sidler, Dominik and Ruggenthaler, Michael and Rubio, Angel and Kowalewski, Markus},
  journal={The {Journal of Physical Chemistry Letters}},
  volume={14},
  number={36},
  pages={8024--8033},
  year={2023},
  publisher={ACS Publications},
    doi = {10.1021/acs.jpclett.3c01842}
}

@article{li2018vacuum,
  title={{{Vacuum Bloch--Siegert}} shift in {{Landau}} polaritons with ultra-high cooperativity},
  author={Li, Xinwei and Bamba, Motoaki and Zhang, Qi and Fallahi, Saeed and Gardner, Geoff C and Gao, Weilu and Lou, Minhan and Yoshioka, Katsumasa and Manfra, Michael J and Kono, Junichiro},
  journal={Nature Photonics},
  volume={12},
  number={6},
  pages={324--329},
  year={2018},
  publisher={Nature Publishing Group UK London},
    doi = {10.1038/s41566-018-0153-0}
}

@article{curtis2019cavity,
  title={Cavity quantum {{E}}liashberg enhancement of superconductivity},
  author={Curtis, Jonathan B and Raines, Zachary M and Allocca, Andrew A and Hafezi, Mohammad and Galitski, Victor M},
  journal={Physical review letters},
  volume={122},
  number={16},
  pages={167002},
  year={2019},
  publisher={APS},
    doi = {10.1103/PhysRevLett.122.167002}
}

@article{mornhinweg2021tailored,
  title={Tailored subcycle nonlinearities of ultrastrong light-matter coupling},
  author={Mornhinweg, Joshua and Halbhuber, M and Ciuti, C and Bougeard, Dominique and Huber, Rupert and Lange, Christoph},
  journal={Physical Review Letters},
  volume={126},
  number={17},
  pages={177404},
  year={2021},
  publisher={APS},
    doi = {10.1103/PhysRevLett.126.177404}
}

@article{juraschek2021cavity,
  title={Cavity control of nonlinear phononics},
  author={Juraschek, Dominik M and Neuman, Tom{\'a}{\v{s}} and Flick, Johannes and Narang, Prineha},
  journal={Physical Review Research},
  volume={3},
  number={3},
  pages={L032046},
  year={2021},
    doi = {10.1103/PhysRevResearch.3.L032046},
  publisher={APS}
}

@article{ruggenthaler.sidler.ea_2023,
  title = {Understanding {{Polaritonic Chemistry}} from {{Ab Initio Quantum Electrodynamics}}},
  author = {Ruggenthaler, Michael and Sidler, Dominik and Rubio, Angel},
  year = {2023},
  month = oct,
  journal = {Chem. Rev.},
  volume = {123},
  number = {19},
  pages = {11191--11229},
  publisher = {American Chemical Society},
  issn = {0009-2665},
  doi = {10.1021/acs.chemrev.2c00788},
}

@article{deMelo2016,
  title = {Unified theory of quantized electrons, phonons, and photons out of equilibrium: A simplified ab initio approach based on the generalized {Baym-Kadanoff} ansatz},
  author = {de Melo, Pedro Miguel M. C. and Marini, Andrea},
  journal = {Phys. Rev. B},
  volume = {93},
  issue = {15},
  pages = {155102},
  numpages = {22},
  year = {2016},
  month = {Apr},
  publisher = {American Physical Society},
  doi = {10.1103/PhysRevB.93.155102},
  url = {https://link.aps.org/doi/10.1103/PhysRevB.93.155102}
}

@article{lu.shin.ea_2025,
  title = {Cavity Engineering of Solid-State Materials without External Driving},
  author = {Lu, I.-Te and Shin, Dongbin and Svendsen, Mark Kamper and Latini, Simone and Hübener, Hannes and Ruggenthaler, Michael and Rubio, Angel},
  year = {2025},
  month = jun,
  journal = {Adv. Opt. Photon.},
  volume = {17},
  number = {2},
  pages = {441--525},
  publisher = {Optica Publishing Group},
  issn = {1943-8206},
  doi = {10/g9mjjd},
  urldate = {2025-05-28},
  copyright = {© 2025 Optica Publishing Group},
  langid = {english}
}

@article{ruggenthaler.tancogne-dejean.ea_2018,
  title = {From a Quantum-Electrodynamical Light–Matter Description to Novel Spectroscopies},
  author = {Ruggenthaler, Michael and {Tancogne-Dejean}, Nicolas and Flick, Johannes and Appel, Heiko and Rubio, Angel},
  year = {2018},
  month = mar,
  journal = {Nat Rev Chem},
  volume = {2},
  number = {3},
  pages = {1--16},
  publisher = {Nature Publishing Group},
  issn = {2397-3358},
  doi = {10/gc8stt},
}

@article{sentef.ruggenthaler.ea_2018,
  title = {Cavity Quantum-Electrodynamical Polaritonically Enhanced Electron-Phonon Coupling and Its Influence on Superconductivity},
  author = {Sentef, M. A. and Ruggenthaler, M. and Rubio, A.},
  year = {2018},
  month = nov,
  journal = {Science Advances},
  volume = {4},
  number = {11},
  pages = {eaau6969},
  publisher = {American Association for the Advancement of Science},
  doi = {10/gfq32g},
}

@article{sortino2025atomic,
  title={Atomic-layer assembly of ultrathin optical cavities in van der {{W}}aals heterostructure metasurfaces},
  author={Sortino, Luca and Biechteler, Jonas and Lafeta, Lucas and K{\"u}hner, Lucca and Hartschuh, Achim and Menezes, Leonardo de S and Maier, Stefan A and Tittl, Andreas},
  journal={Nature Photonics},
volume={19},
  pages={825--832},
  year={2025},
  publisher={Nature Publishing Group},
    doi  = {10.1038/s41566-025-01675-4}
}

@article{mciver2020light,
  title={Light-induced anomalous {Hall} effect in graphene},
  author={McIver, James W and Schulte, Benedikt and Stein, F-U and Matsuyama, Toru and Jotzu, Gregor and Meier, Guido and Cavalleri, Andrea},
  journal={Nature physics},
  volume={16},
  number={1},
  pages={38--41},
  year={2020},
    doi = {10.1038/s41567-019-0698-y},
  publisher={Nature Publishing Group UK London}
}

@article{keimer2015quantum,
  title={From quantum matter to high-temperature superconductivity in copper oxides},
  author={Keimer, Bernhard and Kivelson, Steven A and Norman, Michael R and Uchida, Shinichi and Zaanen, J},
  journal={Nature},
  volume={518},
  number={7538},
  pages={179--186},
  year={2015},
  publisher={Nature Publishing Group UK London},
    doi = {10.1038/nature14165}
}

@article{scalapino2012common,
  title={A common thread: The pairing interaction for unconventional superconductors},
  author={Scalapino, Douglas J},
  journal={Reviews of Modern Physics},
  volume={84},
  number={4},
  pages={1383--1417},
  year={2012},
  publisher={APS},
    doi = {10.1103/RevModPhys.84.1383}
}

@article{buzzi2020photomolecular,
  title={Photomolecular high-temperature superconductivity},
  author={Buzzi, M and Nicoletti, D and Fechner, M and Tancogne-Dejean, N and Sentef, MA and Georges, A and Biesner, T and Uykur, E and Dressel, M and Henderson, A and others},
  journal={Physical Review X},
  volume={10},
  number={3},
  pages={031028},
  year={2020},
  publisher={APS},
    doi = {10.1103/PhysRevX.10.031028}
}

@article{herzig2024high,
  title={High-quality nanocavities through multimodal confinement of hyperbolic polaritons in hexagonal boron nitride},
  author={Herzig Sheinfux, Hanan and Orsini, Lorenzo and Jung, Minwoo and Torre, Iacopo and Ceccanti, Matteo and Marconi, Simone and Maniyara, Rinu and Barcons Ruiz, David and H{\"o}tger, Alexander and Bertini, Ricardo and others},
  journal={Nature Materials},
  volume={23},
  number={4},
  pages={499--505},
  year={2024},
  publisher={Nature Publishing Group UK London},
    doi = {10.1038/s41563-023-01785-w}
}

@article{bacciconi2025theory,
  title={Theory of fractional quantum {Hall} liquids coupled to quantum light and emergent graviton-polaritons},
  author={Bacciconi, Zeno and Xavier, Hernan B and Carusotto, Iacopo and Chanda, Titas and Dalmonte, Marcello},
  journal={Physical Review X},
  volume={15},
  number={2},
  pages={021027},
    doi = {10.1103/PhysRevX.15.021027},
  year={2025},
  publisher={APS}
}

@article{liu2025modifying,
  title={Modifying electronic and structural properties of {2D} van der {{W}}aals materials via cavity quantum vacuum fluctuations: a first-principles {{QEDFT}} study},
  author={Liu, Hang and Latini, Simone and Lu, I-Te and Shin, Dongbin and Rubio, Angel},
  journal={Optical Materials Express},
  volume={15},
  number={9},
  pages={2105--2118},
  year={2025},
  publisher={Optica Publishing Group},
    doi = {10.1364/OME.568454}
}

@article{koo2023tunable,
  title={Tunable interlayer excitons and switchable interlayer trions via dynamic near-field cavity},
  author={Koo, Yeonjeong and Lee, Hyeongwoo and Ivanova, Tatiana and Kefayati, Ali and Perebeinos, Vasili and Khestanova, Ekaterina and Kravtsov, Vasily and Park, Kyoung-Duck},
  journal={Light: Science \& Applications},
  volume={12},
  number={1},
  pages={59},
  year={2023},
  publisher={Nature Publishing Group UK London},
    doi = {10.1038/s41377-023-01087-5}
}

@article{park2019tip,
  title={Tip-enhanced strong coupling spectroscopy, imaging, and control of a single quantum emitter},
  author={Park, Kyoung-Duck and May, Molly A and Leng, Haixu and Wang, Jiarong and Kropp, Jaron A and Gougousi, Theodosia and Pelton, Matthew and Raschke, Markus B},
  journal={Science advances},
  volume={5},
  number={7},
  pages={eaav5931},
  year={2019},
  publisher={American Association for the Advancement of Science},
    doi = {10.1126/sciadv.aav5931}
}

@article{LatiniSTO2021, 
year = {2021}, 
title = {{The ferroelectric photo ground state of {SrTiO}$_3$: Cavity materials engineering.}}, 
author = {Latini, Simone and Shin, Dongbin and Sato, Shunsuke A and Schäfer, Christian and Giovannini, Umberto De and Hübener, Hannes and Rubio, Angel}, 
journal = {Proceedings of the National Academy of Sciences of the United States of America}, 
issn = {0027-8424}, 
doi = {10.1073/pnas.2105618118}, 
pmid = {34315818}, 
pmcid = {PMC8346861}, 
eprint = {2101.11313}, 
pages = {e2105618118}, 
number = {31}, 
volume = {118}
}

@article{zhang2018photonic,
  title={Photonic-crystal exciton-polaritons in monolayer semiconductors},
  author={Zhang, Long and Gogna, Rahul and Burg, Will and Tutuc, Emanuel and Deng, Hui},
  journal={Nature communications},
  volume={9},
  number={1},
  pages={713},
  year={2018},
  publisher={Nature Publishing Group UK London},
    doi = {10.1038/s41467-018-03188-x}
}

@article{huang2023tunable,
  title={Tunable bound states in the continuum in a reconfigurable terahertz metamaterial},
  author={Huang, Yuwei and Kaj, Kelson and Chen, Chunxu and Yang, Zhiwei and Averitt, Richard D and Zhang, Xin},
  journal={Advanced Optical Materials},
  volume={11},
  number={19},
  pages={2300559},
  year={2023},
  publisher={Wiley Online Library},
    doi = {10.1002/adom.202300559}
}

@article{tay2025multimode,
  title={Multimode ultrastrong coupling in three-dimensional photonic-crystal cavities},
  author={Tay, Fuyang and Mojibpour, Ali and Sanders, Stephen and Liang, Shuang and Xu, Hongjing and Gardner, Geoff C and Baydin, Andrey and Manfra, Michael J and Alabastri, Alessandro and Hagenm{\"u}ller, David and others},
  journal={Nature Communications},
  volume={16},
  number={1},
  pages={3603},
  year={2025},
  publisher={Nature Publishing Group UK London},
    doi = {10.1038/s41467-025-58835-x}
}

@article{forn2019ultrastrong,
  title={Ultrastrong coupling regimes of light-matter interaction},
  author={Forn-D{\'\i}az, P and Lamata, L and Rico, E and Kono, J and Solano, E},
  journal={Reviews of Modern Physics},
  volume={91},
  number={2},
  pages={025005},
  year={2019},
  publisher={APS}, 
    doi = {10.1103/RevModPhys.91.025005}
}

@article{sarkar2025sub,
  title={Sub-wavelength optical lattice in {2D} materials},
  author={Sarkar, Supratik and Mehrabad, Mahmoud Jalali and Su{\'a}rez-Forero, Daniel G and Gu, Liuxin and Flower, Christopher J and Xu, Lida and Watanabe, Kenji and Taniguchi, Takashi and Park, Suji and Jang, Houk and others},
  journal={Science Advances},
  volume={11},
  number={13},
  pages={eadv2023},
  year={2025},
  publisher={American Association for the Advancement of Science},
 doi = {10.1126/sciadv.adv2023}
}

@article{blais2021circuit,
  title={Circuit quantum electrodynamics},
  author={Blais, Alexandre and Grimsmo, Arne L and Girvin, Steven M and Wallraff, Andreas},
  journal={Reviews of Modern Physics},
  volume={93},
  number={2},
  pages={025005},
  year={2021},
  publisher={APS},
    doi = {10.1103/RevModPhys.93.025005}
}

@article{enknergraziottofraction,
  title={Tunable vacuum-field control of fractional and integer quantum {Hall} phases},
  author={Enkner, Josefine and Graziotto, Lorenzo and Bori{\c{c}}i, Dalin and Appugliese, Felice and Reichl, Christian and Scalari, Giacomo and Regnault, Nicolas and Wegscheider, Werner and Ciuti, Cristiano and Faist, J{\'e}r{\^o}me},
  journal={Nature},
  volume={641},
    doi = {10.1038/s41586-025-08894-3},
  number={8064},
  pages={884},
  year={2025}
}

@article{arwas.ciuti_2023,
  title = {Quantum Electron Transport Controlled by Cavity Vacuum Fields},
  author = {Arwas, Geva and Ciuti, Cristiano},
  year = {2023},
  month = jan,
  journal = {Physical Review B},
  volume = {107},
  number = {4},
  pages = {045425},
  publisher = {American Physical Society},
  doi = {10.1103/physrevb.107.045425},
}

@article{rokaj.penz.ea_2022,
  title = {Polaritonic {{Hofstadter}} Butterfly and Cavity Control of the Quantized {Hall} Conductance},
  author = {Rokaj, Vasil and Penz, Markus and Sentef, Michael A. and Ruggenthaler, Michael and Rubio, Angel},
  year = {2022},
  month = may,
  journal = {Physical Review B},
  volume = {105},
  number = {20},
  pages = {205424},
  publisher = {American Physical Society},
  doi = {10.1103/physrevb.105.205424},
}

@article{schlawin.cavalleri.ea_2019,
  title = {Cavity-{{Mediated Electron-Photon Superconductivity}}},
  author = {Schlawin, Frank and Cavalleri, Andrea and Jaksch, Dieter},
  year = {2019},
  month = apr,
  journal = {Physical Review Letters},
  volume = {122},
  number = {13},
  pages = {133602},
  publisher = {American Physical Society},
  doi = {10.1103/physrevlett.122.133602},
}

@article{keller2017few,
  title={Few-electron ultrastrong light-matter coupling at 300 {GHz} with nanogap hybrid {LC} microcavities},
  author={Keller, Janine and Scalari, Giacomo and Cibella, Sara and Maissen, Curdin and Appugliese, Felice and Giovine, Ennio and Leoni, Roberto and Beck, Mattias and Faist, J{\'e}r{\^o}me},
  journal={Nano letters},
  volume={17},
  number={12},
  pages={7410--7415},
  year={2017},
  publisher={ACS Publications},
    doi = {10.1021/acs.nanolett.7b03228}
}

@article{nataf2010no,
  title={No-go theorem for superradiant quantum phase transitions in cavity {QED} and counter-example in circuit {QED}},
  author={Nataf, Pierre and Ciuti, Cristiano},
  journal={Nature communications},
  volume={1},
  number={1},
  pages={72},
  year={2010},
  publisher={Nature Publishing Group UK London},
    doi={10.1038/ncomms1069}
}

@article{superradiant_Mazza,
  title = {Superradiant Quantum Materials},
  author = {Mazza, Giacomo and Georges, Antoine},
  journal = {Phys. Rev. Lett.},
  volume = {122},
  issue = {1},
  pages = {017401},
  numpages = {6},
  year = {2019},
  month = {Jan},
  publisher = {American Physical Society},
  doi = {10.1103/PhysRevLett.122.017401},
  url = {https://link.aps.org/doi/10.1103/PhysRevLett.122.017401}
}

@article{masuki.ashida_2024,
  title = {Cavity Moiré Materials: {{Controlling}} Magnetic Frustration with Quantum Light-Matter Interaction},
  shorttitle = {Cavity Moir\textbackslash 'e Materials},
  author = {Masuki, Kanta and Ashida, Yuto},
  year = {2024},
  month = may,
  journal = {Physical Review B},
  volume = {109},
  number = {19},
  pages = {195173},
  publisher = {American Physical Society},
  doi = {10.1103/physrevb.109.195173},
}

@article{gao2024giant,
  title={Giant chiral magnetoelectric oscillations in a van der {Waals} multiferroic},
  author={Gao, Frank Y and Peng, Xinyue and Cheng, Xinle and Vi{\~n}as Bostr{\"o}m, Emil and Kim, Dong Seob and Jain, Ravish K and Vishnu, Deepak and Raju, Kalaivanan and Sankar, Raman and Lee, Shang-Fan and others},
  journal={Nature},
  volume={632},
  number={8024},
  pages={273--279},
  year={2024},
  publisher={Nature Publishing Group UK London},
    doi = {10.1038/s41586-024-07678-5}
}

@article{Bostrom_Ferromagnetism,
  title = {Equilibrium nonlinear phononics by electric field fluctuations of terahertz cavities},
  author = {Vi\~nas Bostr\"om, Emil and Michael, Marios H. and Eckhardt, Christian and Rubio, Angel},
  journal = {Phys. Rev. Res.},
  volume = {7},
  issue = {3},
  pages = {033163},
  numpages = {7},
  year = {2025},
  month = {Aug},
  publisher = {American Physical Society},
  doi = {10.1103/fqq1-9f52},
  url = {https://link.aps.org/doi/10.1103/fqq1-9f52}
}

@article{masuki.ashida_2023,
  title = {Berry Phase and Topology in Ultrastrongly Coupled Quantum Light-Matter Systems},
  author = {Masuki, Kanta and Ashida, Yuto},
  year = {2023},
  month = may,
  journal = {Physical Review B},
  volume = {107},
  number = {19},
  pages = {195104},
  publisher = {American Physical Society},
  doi = {10.1103/physrevb.107.195104},
}

@article{sentef.li.ea_2020,
  title = {Quantum to Classical Crossover of {{Floquet}} Engineering in Correlated Quantum Systems},
  author = {Sentef, Michael A. and Li, Jiajun and Künzel, Fabian and Eckstein, Martin},
  year = {2020},
  month = jul,
  journal = {Physical Review Research},
  volume = {2},
  number = {3},
  pages = {033033},
  publisher = {American Physical Society},
  doi = {10/gtpxq7},
}

@article{li.eckstein_2020,
  title = {Manipulating Intertwined Orders in Solids with Quantum Light},
  author = {Li, Jiajun and Eckstein, Martin},
  year = {2020},
  month = nov,
  journal = {Physical Review Letters},
  volume = {125},
  number = {21},
  pages = {217402},
  publisher = {American Physical Society},
  doi = {10.1103/physrevlett.125.217402},
}

@article{virtual2017, 
    title={Virtual photons in the ground state of a dissipative system},
  author={De Liberato, Simone},
  journal={Nature Communications},
  volume={8},
  number={1},
  pages={1465},
  year={2017},
  publisher={Nature Publishing Group UK London},
    doi = {10.1038/s41467-017-01504-5}
}

@article{basov2025polaritonic,
  title={Polaritonic quantum matter},
  author={Basov, DN and Asenjo-Garcia, Ana and Schuck, P James and Zhu, Xiaoyang and Rubio, Angel and Cavalleri, Andrea and Delor, Milan and Fogler, Michael M and Liu, Mengkun},
  journal={Nanophotonics},
  number={0},
  year={2025},
  publisher={De Gruyter}, 
    doi = {10.1515/nanoph-2025-0001}
}

@article{thomas2021large,
  title={Large enhancement of ferromagnetism under a collective strong coupling of {YBCO} nanoparticles},
  author={Thomas, Anoop and Devaux, Eloise and Nagarajan, Kalaivanan and Rogez, Guillaume and Seidel, Marcus and Richard, Fanny and Genet, Cyriaque and Drillon, Marc and Ebbesen, Thomas W},
  journal={Nano letters},
  volume={21},
  number={10},
    doi = {10.1021/acs.nanolett.1c00973},
  pages={4365--4370},
  year={2021},
  publisher={ACS Publications}
}

@article{jarc_fausti_2023,
  title={Cavity-mediated thermal control of metal-to-insulator transition in {1T-TaS}$_2$},
  author={Jarc, Giacomo and Mathengattil, Shahla Yasmin and Montanaro, Angela and Giusti, Francesca and Rigoni, Enrico Maria and Sergo, Rudi and Fassioli, Francesca and Winnerl, Stephan and Dal Zilio, Simone and Mihailovic, Dragan and others},
  journal={Nature},
  volume={622},
  number={7983},
  pages={487--492},
  year={2023},
  publisher={Nature Publishing Group UK London},
    doi = {10.1038/s41586-023-06596-2}
}

@article{fassioli_fausti_2024,
  title = {Controlling radiative heat flow through cavity electrodynamics},
  author = {Fassioli, Francesca and Faist, Jerome and Eckstein, Martin and Fausti, Daniele},
  journal = {Phys. Rev. B},
  volume = {111},
  issue = {16},
  pages = {165425},
  numpages = {6},
  year = {2025},
  month = {Apr},
  publisher = {American Physical Society},
  doi = {10.1103/PhysRevB.111.165425},
  url = {https://link.aps.org/doi/10.1103/PhysRevB.111.165425}
}

@article{cho2023directional,
  title={Directional radiative cooling via exceptional epsilon-based microcavities},
  author={Cho, Jin-Woo and Lee, Yun-Jo and Kim, Jae-Hyun and Hu, Run and Lee, Eungkyu and Kim, Sun-Kyung},
  journal={{ACS} nano},
  volume={17},
  number={11},
  pages={10442--10451},
  year={2023},
  publisher={ACS Publications}, 
    doi = {10.1021/acsnano.3c01184}
}

@article{pannir2025blackbody,
  title={Blackbody radiation and thermal effects on chemical reactions and phase transitions in cavities},
  author={Pannir-Sivajothi, Sindhana and Yuen-Zhou, Joel},
  journal={ACS nano},
  volume={19},
  number={10},
  pages={9896--9905},
  year={2025},
    doi= {10.1021/acsnano.4c14590},
  publisher={ACS Publications}
}

@article{inoue2015realization,
  title={Realization of narrowband thermal emission with optical nanostructures},
  author={Inoue, Takuya and De Zoysa, Menaka and Asano, Takashi and Noda, Susumu},
  journal={Optica},
  volume={2},
  number={1},
  pages={27--35},
  year={2015},
    doi = {10.1364/OPTICA.2.000027},
  publisher={Optical Society of America}
}

@article{jarc_fausti_2024,
  title={Multimode vibrational coupling across the insulator-to-metal transition in {1T-TaS}$_2$ in {THz} cavities},
  author={Jarc, Giacomo and Mathengattil, Shahla Yasmin and Montanaro, Angela and Rigoni, Enrico Maria and Dal Zilio, Simone and Fausti, Daniele},
  journal={The Journal of Chemical Physics},
  volume={161},
  number={15},
  year={2024},
  publisher={AIP Publishing},
    doi = {10.1063/5.0231533}
}

@article{flores_piazza_2025,
  title={Nonthermal electron-photon steady states in open cavity quantum materials},
  author={Flores-Calder{\'o}n, R and Islam, Md Mursalin and Pini, Michele and Piazza, Francesco},
  journal={Physical Review Research},
  volume={7},
  number={1},
  pages={013073},
  year={2025},
  publisher={APS},
    doi = {10.1103/PhysRevResearch.7.013073}
}

@article{CiutiTopology20241D, 
year = {2024}, 
title = {{Electron conductance and many-body marker of a cavity-embedded topological one-dimensional chain}}, 
author = {Nguyen, Danh-Phuong and Arwas, Geva and Ciuti, Cristiano}, 
journal = {Physical Review B}, 
issn = {2469-9950}, 
doi = {10.1103/physrevb.110.195416}, 
pages = {195416}, 
number = {19}, 
volume = {110}
}

@article{QSL_Natcom2021, 
year = {2021}, 
title = {{Cavity-induced quantum spin liquids}}, 
author = {Chiocchetta, Alessio and Kiese, Dominik and Zelle, Carl Philipp and Piazza, Francesco and Diehl, Sebastian}, 
journal = {Nature Communications}, 
doi = {10.1038/s41467-021-26076-3}, 
pmid = {34625551}, 
pmcid = {PMC8501047}, 
pages = {5901}, 
number = {1}, 
volume = {12}
}

@article{hubener2021engineering,
  title={Engineering quantum materials with chiral optical cavities},
  author={H{\"u}bener, Hannes and De Giovannini, Umberto and Sch{\"a}fer, Christian and Andberger, Johan and Ruggenthaler, Michael and Faist, Jerome and Rubio, Angel},
  journal={Nature materials},
  volume={20},
  number={4},
    doi = {10.1038/s41563-020-00801-7},
  pages={438--442},
  year={2021},
  publisher={Nature Publishing Group UK London}
}

@article{wang2019cavity,
  title={Cavity quantum electrodynamical Chern insulator: Towards light-induced quantized anomalous {Hall} effect in graphene},
  author={Wang, Xiao and Ronca, Enrico and Sentef, Michael A},
  journal={Physical Review B},
  volume={99},
  number={23},
  pages={235156},
  year={2019},
  publisher={APS}
}

@article{raja2017coulomb,
  title={Coulomb engineering of the bandgap and excitons in two-dimensional materials},
  author={Raja, Archana and Chaves, Andrey and Yu, Jaeeun and Arefe, Ghidewon and Hill, Heather M and Rigosi, Albert F and Berkelbach, Timothy C and Nagler, Philipp and Sch{\"u}ller, Christian and Korn, Tobias and others},
  journal={Nature communications},
  volume={8},
  number={1},
  pages={15251},
    doi = {10.1038/ncomms15251},
  year={2017},
  publisher={Nature Publishing Group UK London}
}

@article{keimer2017quantum,
  title={The physics of quantum materials},
  author={Keimer, Bernhard and Moore, Joel E.},
  journal={Nature Physics},
  volume={13},
  pages={1045},
    doi = {10.1038/nphys4302},
  year={2017},
  publisher={Nature Publishing Group UK London}
}

@article{ashida2020quantum,
  title={Quantum electrodynamic control of matter: Cavity-enhanced ferroelectric phase transition},
  author={Ashida, Yuto and {\.I}mamo{\u{g}}lu, Ata{\c{c}} and Faist, J{\'e}r{\^o}me and Jaksch, Dieter and Cavalleri, Andrea and Demler, Eugene},
  journal={Physical Review X},
  volume={10},
  number={4},
    doi = {10.1103/PhysRevX.10.041027},
  pages={041027},
  year={2020},
  publisher={APS}
}

@article{curtis2023local,
  title={Local fluctuations in cavity control of ferroelectricity},
  author={Curtis, Jonathan B and Michael, Marios H and Demler, Eugene},
  journal={Physical Review Research},
  volume={5},
  number={4},
    doi = {10.1103/PhysRevResearch.5.043118}, 
  pages={043118},
  year={2023},
  publisher={APS}
}

@article{bustamante2025molecular,
  title={Molecular polariton dynamics in realistic cavities},
  author={Bustamante, Carlos M and Bonaf{\'e}, Franco P and Sukharev, Maxim and Ruggenthaler, Michael and Nitzan, Abraham and Rubio, Angel},
  journal={Journal of Chemical Theory and Computation},
  year={2025},
    doi={10.1021/acs.jctc.5c01318},
  publisher={ACS Publications}
}

@article{svendsen2024ab,
  title={Ab initio calculations of quantum light--matter interactions in general electromagnetic environments},
  author={Svendsen, Mark Kamper and Thygesen, Kristian Sommer and Rubio, Angel and Flick, Johannes},
  journal={Journal of Chemical Theory and Computation},
  volume={20},
  number={2},
  pages={926--936},
  year={2024},
    doi = {10.1021/acs.jctc.3c00967}, 
  publisher={ACS Publications}
}

@article{yoshioka2020onchip, 
  year     = {2020}, 
  title    = {On-chip coherent frequency-domain {THz} spectroscopy for electrical transport}, 
  author   = {Yoshioka, Katsumasa and Kumada, Norio and Muraki, Koji and Hashisaka, Masayuki}, 
  journal  = {Applied Physics Letters}, 
  issn     = {0003-6951}, 
  doi      = {10.1063/5.0024089}, 
  pages    = {161103}, 
  number   = {16}, 
  volume   = {117}
}

@article{jiang2024engineering,
  title={Engineering flat bands in twisted-bilayer graphene away from the magic angle with chiral optical cavities},
  author={Jiang, Cunyuan and Baggioli, Matteo and Jiang, Qing-Dong},
  journal={Physical Review Letters},
  volume={132},
  number={16},
  pages={166901},
  year={2024},
  publisher={APS},
    doi = {10.1103/PhysRevLett.132.166901}
}

@article{eckhardt2022quantum,
  title={Quantum {Floquet} engineering with an exactly solvable tight-binding chain in a cavity},
  author={Eckhardt, Christian J and Passetti, Giacomo and Othman, Moustafa and Karrasch, Christoph and Cavaliere, Fabio and Sentef, Michael A and Kennes, Dante M},
  journal={Communications Physics},
  volume={5},
  number={1},
  pages={122},
  year={2022},
    doi = {10.1038/s42005-022-00880-9},
  publisher={Nature Publishing Group UK London}
}

@article{nguyen2023electron,
  title={Electron-photon {Chern} number in cavity-embedded {2D} moir{\'e} materials},
  author={Nguyen, Danh-Phuong and Arwas, Geva and Lin, Zuzhang and Yao, Wang and Ciuti, Cristiano},
  journal={Physical Review Letters},
  volume={131},
  number={17},
  pages={176602},
  year={2023},
  publisher={APS}, 
    doi = {10.1103/PhysRevLett.131.176602}
}

@article{riek2015direct,
  title={Direct sampling of electric-field vacuum fluctuations},
  author={Riek, Claudius and Seletskiy, Denis V and Moskalenko, Andrey S and Schmidt, JF and Krauspe, Philipp and Eckart, Sebastian and Eggert, Stefan and Burkard, Guido and Leitenstorfer, Alfred},
  journal={Science},
  volume={350},
  number={6259},
  pages={420--423},
  year={2015},
  publisher={American Association for the Advancement of Science},
    doi = {10.1126/science.aac9788}
}

@article{benea2019electric,
  title={Electric field correlation measurements on the electromagnetic vacuum state},
  author={Benea-Chelmus, Ileana-Cristina and Settembrini, Francesca Fabiana and Scalari, Giacomo and Faist, J{\'e}r{\^o}me},
  journal={Nature},
  volume={568},
  number={7751},
  pages={202--206},
  year={2019},
    doi = {10.1038/s41586-019-1083-9},
  publisher={Nature Publishing Group UK London}
}

@article{paravicini2019magneto,
  title={Magneto-transport controlled by {{Landau}} polariton states},
  author={Paravicini-Bagliani, Gian L and Appugliese, Felice and Richter, Eli and Valmorra, Federico and Keller, Janine and Beck, Mattias and Bartolo, Nicola and R{\"o}ssler, Clemens and Ihn, Thomas and Ensslin, Klaus and others},
  journal={Nature Physics},
  volume={15},
  number={2},
  pages={186--190},
  year={2019},
  publisher={Nature Publishing Group UK London},
    doi = {10.1038/s41567-018-0346-y}
}

@article{rokaj2023weakened,
  title={Weakened topological protection of the quantum {Hall} effect in a cavity},
  author={Rokaj, Vasil and Wang, Jie and Sous, John and Penz, Markus and Ruggenthaler, Michael and Rubio, Angel},
  journal={Physical Review Letters},
  volume={131},
  number={19},
  pages={196602},
  year={2023},
  publisher={APS}, 
    doi = {10.1103/PhysRevLett.131.196602}
}

@article{ciuti2021cavity,
  title={Cavity-mediated electron hopping in disordered quantum {Hall} systems},
  author={Ciuti, Cristiano},
  journal={Physical Review B},
  volume={104},
  number={15},
  pages={155307},
  year={2021},
  publisher={APS}, 
    doi = {10.1103/PhysRevB.104.155307}
}

@article{appugliese2022breakdown,
  title={Breakdown of topological protection by cavity vacuum fields in the integer quantum {Hall} effect},
  author={Appugliese, Felice and Enkner, Josefine and Paravicini-Bagliani, Gian Lorenzo and Beck, Mattias and Reichl, Christian and Wegscheider, Werner and Scalari, Giacomo and Ciuti, Cristiano and Faist, J{\'e}r{\^o}me},
  journal={Science},
  volume={375},
  number={6584},
  pages={1030--1034},
  year={2022},
    doi = {10.1126/science.abl5818},
  publisher={American Association for the Advancement of Science}
}

@article{Entanglement2024,
  title = {Entanglement harvesting from electromagnetic quantum fields},
  author = {Lindel, Frieder and Herter, Alexa and Gebhart, Valentin and Faist, J\'er\^ome and Buhmann, Stefan Yoshi},
  journal = {Phys. Rev. A},
  volume = {110},
  issue = {2},
  pages = {022414},
  numpages = {19},
  year = {2024},
  month = {Aug},
  publisher = {American Physical Society},
  doi = {10.1103/PhysRevA.110.022414},
  url = {https://link.aps.org/doi/10.1103/PhysRevA.110.022414}
}

@article{benea2025electro,
  title={Electro-optic sampling of classical and quantum light},
  author={Benea-Chelmus, Ileana-Cristina and Faist, J{\'e}r{\^o}me and Leitenstorfer, Alfred and Moskalenko, Andrey S and Pupeza, Ioachim and Seletskiy, Denis V and Vodopyanov, Konstantin L},
  journal={Optica},
  volume={12},
  number={4},
  pages={546--563},
  year={2025},
  publisher={Optica Publishing Group},
  doi = {10.1364/OPTICA.544333}
}

@article{graziotto2025cavity,
title={Cavity {QED} Control of Quantum {Hall} Stripes},
author={Graziotto, Lorenzo and Enkner, Josefine and Chattopadhyay, Sambuddha and Curtis, Jonathan and Koskas, Ethan and Reichl, Christian and Wegscheider, Werner and Scalari, Giacomo and Demler, Eugene and Faist, J{\'e}r{\^o}me},
  journal={arXiv},
  year={2025},
    eprint = {2502.15490}
}

@article{koulakov1996charge,
  title={Charge density wave in two-dimensional electron liquid in weak magnetic field},
  author={Koulakov, AA and Fogler, MM and Shklovskii, Boris I},
  journal={Physical review letters},
  volume={76},
  number={3},
  pages={499},
  year={1996},
  publisher={APS},
    doi = {10.1103/PhysRevLett.76.499}
}

@article{frisk_kockum_ultrastrong_2019,
    title = {Ultrastrong coupling between light and matter},
    volume = {1},
    copyright = {2019 Springer Nature Limited},
    issn = {2522-5820},
    url = {https://www.nature.com/articles/s42254-018-0006-2},
    doi = {10.1038/s42254-018-0006-2},
    number = {1},
    urldate = {2021-09-22},
    journal = {Nature Reviews Physics},
    author = {Frisk Kockum, Anton and Miranowicz, Adam and De Liberato, Simone and Savasta, Salvatore and Nori, Franco},
    month = jan,
    year = {2019},
    pages = {19--40},
}

@article{suarez-forero_chiral_2024,
	title = {Chiral flat-band optical cavity with atomically thin mirrors},
	volume = {10},
	url = {https://www.science.org/doi/full/10.1126/sciadv.adr5904},
	doi = {10.1126/sciadv.adr5904},
	number = {51},
	urldate = {2024-12-31},
	journal = {Science Advances},
	author = {Suárez-Forero, Daniel G. and Ni, Ruihao and Sarkar, Supratik and Jalali Mehrabad, Mahmoud and Mechtel, Erik and Simonyan, Valery and Grankin, Andrey and Watanabe, Kenji and Taniguchi, Takashi and Park, Suji and Jang, Houk and Hafezi, Mohammad and Zhou, You},
	month = dec,
	year = {2024},
	pages = {eadr5904},
}

@article{eckhardt_theory_2024,
	title = {Theory of resonantly enhanced photo-induced superconductivity},
	volume = {15},
	copyright = {2024 The Author(s)},
	issn = {2041-1723},
	url = {https://www.nature.com/articles/s41467-024-46632-x},
	doi = {10.1038/s41467-024-46632-x},
	number = {1},
	urldate = {2025-08-26},
	journal = {Nature Communications},
	author = {Eckhardt, Christian J. and Chattopadhyay, Sambuddha and Kennes, Dante M. and Demler, Eugene A. and Sentef, Michael A. and Michael, Marios H.},
	month = mar,
	year = {2024},
	pages = {2300},
}

@article{de_la_torre_colloquium_2021,
	title = {Colloquium: {Nonthermal} pathways to ultrafast control in quantum materials},
	volume = {93},
	shorttitle = {Colloquium},
	url = {https://link.aps.org/doi/10.1103/RevModPhys.93.041002},
	doi = {10.1103/RevModPhys.93.041002},
	number = {4},
	urldate = {2025-03-10},
	journal = {Reviews of Modern Physics},
	author = {de la Torre, Alberto and Kennes, Dante M. and Claassen, Martin and Gerber, Simon and McIver, James W. and Sentef, Michael A.},
	month = oct,
	year = {2021},
	pages = {041002},
}

@article{passetti_cavity_2023,
	title = {Cavity light-matter entanglement through quantum fluctuations},
	volume = {131},
	url = {https://link.aps.org/doi/10.1103/PhysRevLett.131.023601},
	doi = {10.1103/PhysRevLett.131.023601},
	number = {2},
	urldate = {2024-10-30},
	journal = {Physical Review Letters},
	author = {Passetti, Giacomo and Eckhardt, Christian J. and Sentef, Michael A. and Kennes, Dante M.},
	month = jul,
	year = {2023},
	pages = {023601},
}

@article{reutzel_nonlinear_2019,
	title = {Nonlinear {Plasmonic} {Photoelectron} {Response} of {Ag}(111)},
	volume = {123},
	issn = {0031-9007, 1079-7114},
	url = {https://link.aps.org/doi/10.1103/PhysRevLett.123.017404},
	doi = {10.1103/PhysRevLett.123.017404},
	number = {1},
	urldate = {2025-08-26},
	journal = {Physical Review Letters},
	author = {Reutzel, Marcel and Li, Andi and Gumhalter, Branko and Petek, Hrvoje},
	month = jul,
	year = {2019},
	pages = {017404},
}

@article{zhou_cavity_2024,
	title = {Cavity {Floquet} engineering},
	volume = {15},
	copyright = {2024 The Author(s)},
	issn = {2041-1723},
	url = {https://www.nature.com/articles/s41467-024-52014-0},
	doi = {10.1038/s41467-024-52014-0},
	number = {1},
	urldate = {2025-08-26},
	journal = {Nature Communications},
	author = {Zhou, Lingxiao and Liu, Bin and Liu, Yuze and Lu, Yang and Li, Qiuyang and Xie, Xin and Lydick, Nathanial and Hao, Ruofan and Liu, Chenxi and Watanabe, Kenji and Taniguchi, Takashi and Chou, Yu-Hsun and Forrest, Stephen R. and Deng, Hui},
	month = sep,
	year = {2024},
	pages = {7782},
}

@article{andolina_amperean_2024,
	title = {Amperean superconductivity cannot be induced by deep subwavelength cavities in a two-dimensional material},
	volume = {109},
	issn = {2469-9950, 2469-9969},
	url = {https://link.aps.org/doi/10.1103/PhysRevB.109.104513},
	doi = {10.1103/PhysRevB.109.104513},
	number = {10},
	urldate = {2024-12-31},
	journal = {Physical Review B},
	author = {Andolina, Gian Marcello and De Pasquale, Antonella and Pellegrino, Francesco Maria Dimitri and Torre, Iacopo and Koppens, Frank H. L. and Polini, Marco},
	month = mar,
	year = {2024},
	pages = {104513},
}

@article{basov_polariton_2021,
	title = {Polariton panorama},
	volume = {10},
	issn = {2192-8614},
	url = {https://www.degruyter.com/document/doi/10.1515/nanoph-2020-0449/html?lang=en&srsltid=AfmBOopevnopuJxKxzzoY3NLA_vnkj8yV3IWzxlQPjG5O0WnZAPBs5zz},
	doi = {10.1515/nanoph-2020-0449},
	number = {1},
	urldate = {2025-03-12},
	journal = {Nanophotonics},
	author = {Basov, D. N. and Asenjo-Garcia, Ana and Schuck, P. James and Zhu, Xiaoyang and Rubio, Angel},
	month = jan,
	year = {2021},
	pages = {549--577},
}

@article{basov_towards_2017,
	title = {Towards properties on demand in quantum materials},
	volume = {16},
	copyright = {2017 Springer Nature Limited},
	issn = {1476-4660},
	url = {https://www.nature.com/articles/nmat5017},
	doi = {10.1038/nmat5017},
	number = {11},
	urldate = {2025-03-10},
	journal = {Nature Materials},
	author = {Basov, D. N. and Averitt, R. D. and Hsieh, D.},
	month = nov,
	year = {2017},
	pages = {1077--1088},
}

@article{bradley_quantum_2024,
	title = {Quantum {Wire} {Coupled} to {Light}},
	volume = {5},
	url = {https://link.aps.org/doi/10.1103/PRXQuantum.5.040338},
	doi = {10.1103/PRXQuantum.5.040338},
	number = {4},
	urldate = {2024-12-31},
	journal = {PRX Quantum},
	author = {Bradley, Victor and Sharma, Kamal and Hafezi, Mohammad and DeGottardi, Wade},
	month = dec,
	year = {2024},
	Publisher = {APS},
	pages = {040338},
}

@article{grunwald2024cavity,
  title = {Cavity Spectroscopy for Strongly Correlated Polaritonic Systems},
  author = {Grunwald, Lukas and Bostr\"om, Emil Vinas and Svendsen, Mark Kamper and Kennes, Dante M. and Rubio, Angel},
  journal = {Phys. Rev. Lett.},
  volume = {134},
  issue = {24},
  pages = {246901},
  numpages = {7},
  year = {2025},
  month = {Jun},
  publisher = {APS},
  doi = {10.1103/1lpw-22np},
  url = {https://link.aps.org/doi/10.1103/1lpw-22np}
}

@article{vinas_bostrom_controlling_2023,
	title = {Controlling the magnetic state of the proximate quantum spin liquid $\alpha$-{RuCl}$_3$ with an optical cavity},
	volume = {9},
	copyright = {2023 The Author(s)},
	issn = {2057-3960},
	url = {https://www.nature.com/articles/s41524-023-01158-6},
	doi = {10.1038/s41524-023-01158-6},
	number = {1},
	urldate = {2024-12-13},
	journal = {npj Computational Materials},
	author = {Viñas Boström, Emil and Sriram, Adithya and Claassen, Martin and Rubio, Angel},
	month = oct,
	year = {2023},
	pages = {1--10},
}

@article{garcia-vidal_manipulating_2021,
	title = {Manipulating matter by strong coupling to vacuum fields},
	volume = {373},
	url = {https://www.science.org/doi/10.1126/science.abd0336},
	doi = {10.1126/science.abd0336},
	number = {6551},
	urldate = {2024-05-28},
	journal = {Science},
	author = {Garcia-Vidal, Francisco J. and Ciuti, Cristiano and Ebbesen, Thomas W.},
	month = jul,
	year = {2021},
	pages = {eabd0336},
}

@article{schlawin_cavity_2022,
	title = {Cavity quantum materials},
	volume = {9},
	url = {https://aip.scitation.org/doi/10.1063/5.0083825},
	doi = {10.1063/5.0083825},
	number = {1},
	urldate = {2022-03-24},
	journal = {Applied Physics Reviews},
	author = {Schlawin, F. and Kennes, D. M. and Sentef, M. A.},
	month = mar,
	year = {2022},
	pages = {011312},
}

\end{document}